\documentclass{aastex}

\def \be{\begin{eqnarray}}
\def \ee{\end{eqnarray}}

\def\thalf{{\textstyle{\frac{1}{2}}}}

\received{}
\revised{}
\accepted{}
\shorttitle{Evolution of PNS with kaon condensates}
\shortauthors{Pons et al.}
\begin{document}

\setlength\baselineskip{15pt}

\title{ EVOLUTION OF PROTO--NEUTRON STARS \\
WITH KAON CONDENSATES}
\author{J.A. PONS$^{1,a}$, J.A. MIRALLES$^{2}$,
M. PRAKASH$^{1}$ and J.M. LATTIMER$^{1}$}
\affil{
$^1$Department of Physics \& Astronomy\\
SUNY at Stony Brook, Stony Brook, NY 11794-3800\\
$^2$Departament  d'Astronomia i Astrof\'{\i}sica \\
Universitat de Val\`{e}ncia, 46100 Burjassot, Val\`{e}ncia, Spain \\
} 
\email{$^{a}$jpons@neutrino.ess.sunysb.edu \\
}

\begin{abstract}

We present simulations of the evolution of a proto-neutron
star in which kaon-condensed matter might exist, including the effects
of finite temperature and trapped neutrinos. The phase transition from
pure nucleonic matter to the kaon condensate phase is described using
Gibbs' rules for phase equilibrium, which permit the existence of a
mixed phase.  A general property of neutron stars containing kaon
condensates, as well as other forms of strangeness, is that the
maximum mass for cold, neutrino-free matter can be less than the
maximum mass for matter containing trapped neutrinos or which has a
finite entropy. A proto-neutron star formed with a baryon mass
exceeding that of the maximum mass of cold, neutrino-free matter is
therefore metastable, that is, it will collapse to a black hole at
some time during the Kelvin-Helmholtz cooling stage.  

The effects of kaon condensation on metastable stars are dramatic.  In
these cases, the neutrino signal from a hypothetical galactic
supernova (distance $\sim8.5$ kpc) will stop suddenly, generally at a
level above the background in the SuperK and SNO detectors, which have
low energy thresholds and backgrounds. This is in contrast to the case
of a stable star, for which the signal exponentially decays,
eventually disappearing into the background.  We find the lifetimes of
kaon-condensed metastable stars to be restricted to the range 40--70 s
and weakly dependent on the proto-neutron star mass, in sharp contrast
to the significantly larger mass dependence and range (1--100 s) of
hyperon-rich metastable stars.

We find that a unique signature for kaon condensation will be
difficult to identify.  The formation of the kaon condensate is
delayed until the final stages of the Kelvin-Helmholtz epoch, when the
neutrino luminosity is relatively small.  In stable stars, modulations
of the neutrino signal caused by the appearance of the condensate will
therefore be too small to be clearly distinguished with current
detectors, despite the presence of a first order phase transition in
the core.  In metastable stars, the sudden cessation in the neutrino
signal occurs whether it is caused by kaon condensation, hyperons or
quarks.  However, if the lifetime of the metastable star is less than
about 30 s, we find it is not likely to be due to kaon condensation.

\end{abstract}



\section{INTRODUCTION}

Proto-neutron stars (PNSs) are formed in the aftermath of
gravitational collapse supernova, the end state of stars more massive
than about 8 M$_\odot$.  These objects are prodigious emitters of
neutrinos of all types which, if detected in terrestrial detectors,
could reveal details of the supernova mechanism and the properties and
composition of dense matter.  The observation of neutrinos from SN 1987A
\citep{Bio87,Hir87} confirmed the standard
scenario for the early evolution of PNSs \citep{BL86}.
At the beginning, this PNS is very hot and lepton-rich and, after a
typical time of several tens of seconds, the star becomes deleptonized
and cold: a neutron star has been formed.  The detection of neutrinos
radiating from the PNS surface is an unequivocal signal of the
formation of this kind of object, since the direct collapse of the
iron core into a black hole that might occur, perhaps in relatively
extremely massive stars due to accretion, would result in a neutrino
signal of rather short duration \citep{Bur88}.

One of the chief objectives in modeling PNSs is to determine their
internal compositions.  Many simulations of dense matter predict the
appearance of strange matter, in the guise of hyperons, a kaon
condensate, or quark matter at supernuclear density 
\citep[and references therein]{Pra97}.  An important question
is whether or not neutrino observations from a supernova could reveal
the presence of such matter.  One interesting possibility is that the
existence of strange matter in neutron stars makes a sufficiently
massive PNS metastable, so that after a delay of 10--100 seconds, the
PNS collapses into a black hole 
\citep{TPL94,Bro94,BB94,PCL95,Gle95,KJ95,ELP96}.
Such an event might be
straightforward to observe as an abrupt cessation of neutrinos when
the instability is triggered.  Once the star becomes unstable, the
collapse to a black hole proceeds on a time-scale much shorter than
the diffusion time scale, and the neutrinos still trapped in the inner
regions cannot escape \citep{BST96,Bau96}.

A previous paper \cite[paper I hereafter]{Pon99} presented
calculations of the evolution of PNSs, studying the sensitivity of the
results to the initial model, the total mass, the underlying equation
of state (EOS) and the possible presence of hyperons.  PNS simulations
of stars with hyperons, not including hyperon contributions to the
opacity, were earlier performed by \citet{KJ95}.  Paper I showed
that the major effect on the neutrino signal before the onset of any
possible metastability is the PNS mass: larger masses give rise to
larger luminosities and generally higher average emitted neutrino
energies.  In addition, it was found that mass windows for metastable 
models could be as large as 0.3 M$_\odot$, ranging from baryon masses 
$M_B=1.7$ M$_\odot$ to 2.0  M$_\odot$.  The lifetimes of hyperonic
metastable stars decrease with stellar mass and range from a few to
longer than 100 s.  The detection of neutrinos from SN 1987A over a
timescale of 10-15 s is thus consistent with either the formation of a
stable PNS or a metastable PNS containing hyperons, as long as its
mass was less than about 0.1 M$_\odot$ below the maximum mass for
cold, catalyzed hyperonic matter.  Larger PNS masses would lead to a
collapse to a black hole on a timescale shorter than that observed.

A similar situation could be encountered if the EOS allowed the
presence of other forms of ``exotic'' matter, manifested in the form
of a Bose condensate (of pions or kaons) or quarks 
\citep{PCL95,ELP96,SPL00}. As suggested some years ago \citep{KN86},
kaons obtain an effective mass which decreases with density in dense
matter. As a consequence, the ground state of hadronic matter at high
density might contain a kaon condensate  \citep{Bro92,MT92,TPL94,Mar94}, 
which would result in a much softer EOS \citep[hereafter paper II]
{TY99,GS98,Pon00}. However, a high lepton
fraction, which exists in the early PNS evolution, suppresses the
formation of kaons just as hyperon formation or the appearance of
quarks is impeded \citep{PCL95,Pra97}.
The appearance of kaons after deleptonization
produces a softening of the EOS, which can destabilize a sufficiently
massive star, causing a collapse into a black hole.

In this paper we concentrate on models obtained in the context of kaon
condensation and explore their minimal and maximal effects.  One
important difference between models containing hyperons and those
containing a kaon condensate stems from the first order phase
transition that often exists in the latter case.  For some models of
kaon-nucleon interactions, mixed phase regions can be present.
Previous works \citep[paper II]{Gle92,PCL95,GS99} have emphasized that
satisfying the Gibbs' phase rules for equilibrium in situations when
more than one conserved charge (in our case, baryon number and
electric charge, lepton number being constrained by beta equilibrium)
is present can radically change the pressure-density relation compared
to the case in which a Maxwell construction is employed.  
\citet{KPE95,SM95} have
emphasized that the threshold density for the appearance of strange
particles, hyperons and kaons, are sensitive to the rather poorly known
interactions in dense matter. In models in which hyperons appear at
lower densities than kaons, the role of kaons can be relatively small.
On the other hand, should it turn out that the interactions of the
charged hyperons are relatively repulsive, only neutral hyperons may
be present and kaons can play a dominant role.

Although the possibility of PNS metastability was first discovered
in the context of kaon condensation in neutron star matter \citep{TPL94},
a full simulation of a PNS
evolution with a kaon condensate has not been performed so far.
\citet{BST96} have previously studied the hydrodynamical
collapse of metastable PNSs containing kaon condensates.  They were,
however, mostly interested in the dynamical collapse to a black hole,
not in the PNS evolution.  They employed the kaon EOS from \citet{TPL94}
in which the transition from normal matter
to the phase containing kaons was described by a Maxwell construction.
They also rescaled the neutrino mean free paths by a factor of a
thousand to accelerate the quasi-static part (i.e., the
Kelvin-Helmholtz phase) of the evolution, in order to save computation
time.  This produces little effect on the collapse to a black hole,
since neutrinos are essentially trapped in the short time of collapse.
However, their results for the Kelvin-Helmholtz stage can only be
considered as rough estimates, because of this over-simplification.

In this paper, we perform the first evolutionary simulations of PNSs
with kaon condensates during the Kelvin-Helmholtz epoch by using
Henyey-like evolution and flux-limited diffusion modules as described
in paper I.  As in paper I, we ignore the effects of accretion, which
are generally believed to be small after a successful supernova
explosion, expected to occur within 1 s of core bounce.  We use a
field-theoretical EOS that includes kaon condensation (paper II) using
boson exchange interactions and employ Gibbs' phase rules for
multicomponent matter in which baryon number, electric charge, and
lepton number are conserved during the evolution.  We use the
opacities developed by \citet{RPL98,Red99} and \citet{BS98,BS99} in
the nucleon sector and account for the modifications that arise from
the presence of kaon-condensed matter \citep{Tho95,JM98,Red98}.

Our presentation is organized as follows. \S2 provides an outline of
the basic features of the EOS of matter containing a kaon
condensate. In \S3 we describe the thermodynamical properties of
kaon-condensed matter at finite temperature and with varying amounts
of leptons likely to be encountered in the evolution of a PNS.  \S4
contains a description of the neutrino opacities in kaon condensed
matter.  Results from numerical simulations of the Kelvin--Helmholtz
phase of the evolution of a PNS with a kaon condensate are presented
in \S5. Here we also explore the sensitivity of the results to the initial
mass and consequences associated with metastable PNSs which collapse
into a black hole. The detectability of black hole formation in the
Super-Kamiokande (SuperK) and SNO detectors from possible future galactic
supernovae are investigated in \S6.  Our conclusions are
given in \S7.

\section{EQUATION OF STATE OF KAON CONDENSED  MATTER}

The EOS of kaon-condensed matter including the effects of trapped
neutrinos and finite temperature for several different
field-theoretical models for both the nucleon-nucleon and kaon-nucleon
interactions was studied in paper II.  Here we summarize the basic
relations required for the computation of the EOS.  Since our purpose
in this paper is to explore whether a discriminating neutrino signal
would be observed from kaon condensed matter, we present results for
cases in which a first order phase transition occurs at zero
temperature.

To model the baryonic phase, we use a field-theoretical description in
which baryons interact via the exchange of $\sigma,\omega$, and $\rho$
mesons. Including leptons, the Lagrangian density is given by \citep{SW86},
\begin{eqnarray}
L &=& L_H +L_\ell \nonumber \\
  &=& \sum_{B} \overline{B}(-i\gamma^{\mu}\partial_{\mu}-g_{\omega B}
\gamma^{\mu}\omega_\mu
-g_{\rho B}\gamma^{\mu}{\bf{b}}_{\mu}\cdot{\bf t}-M^*)B \nonumber \\
&-& \frac{1}{4}W_{\mu\nu}W^{\mu\nu}+\frac{1}{2}m_{\omega}^2\omega_{\mu}\omega^
{\mu} - \frac{1}{4}{\bf B_{\mu\nu}}{\bf
B^{\mu\nu}}+\frac{1}{2}m_{\rho}^2 b_{\mu}b^{\mu} \nonumber \\
&+& \frac{1}{2}\partial_{\mu}\sigma\partial^{\mu}\sigma -\frac{1}{2}
m_{\sigma}^2\sigma^2-U(\sigma)
+ \sum_{l}\overline{l}(-i\gamma^{\mu}\partial_{\mu}-m_l)l \,.
\end{eqnarray}
Here, $B$ are the Dirac spinors for baryons, $\bf t$ is the isospin
operator, and the $g$s and $m$s are the couplings and masses of the
mesons.  The sums include baryons $B=n,p$, and leptons, $l = e^-$,
$\mu^-$ and $\nu_i$. The field strength tensors for the $\omega$ and
$\rho$ mesons are $W_{\mu\nu} =
\partial_\mu\omega_\nu-\partial_\nu\omega_\mu$ and ${\bf B}_{\mu\nu} =
\partial_\mu{\bf b}_\nu-\partial_\nu{\bf b}_\mu$, respectively.  The
potential $U(\sigma)$ represents the self-interactions of the scalar
field and is taken to be of the form \citep{BB77}
\begin{eqnarray}
U(\sigma) =  \frac{1}{3}bm_B(g_{\sigma N}\sigma)^3 + \frac{1}{4}c(g_{\sigma
N}\sigma)^4\,,
\end{eqnarray}
where $m_B$ is the bare nucleon mass.

In the mean field approximation the thermodynamic potential per unit
volume is \citep{SW86}
\begin{equation}
\frac{\Omega_{\cal N}}{V}=\thalf m_{\sigma}^2\sigma^2+U(\sigma)
-\thalf m_{\omega}^2\omega_0^2-\thalf m_{\rho}^2b_0^2
- 2T\sum_{n,p}\int\frac{d^3k}{(2\pi)^3} \,\ln\left(1+{\rm e}
^{-\beta(E^*-\nu_{n,p})}\right)\;.
\label{hyp2}
\end{equation}
Here the inverse temperature is denoted by $\beta=1/T$ and
$E^*=\sqrt{k^2+M^{*2}}$ with $M^* = m_B - g_{\sigma} \sigma$ denoting
the nucleon effective mass.  The chemical potentials are given by
\begin{equation}
\mu_p=\nu_p+g_{\omega}\omega_0+\thalf g_{\rho}b_0\quad;\quad
\mu_n=\nu_n+g_{\omega}\omega_0-\thalf g_{\rho}b_0\;.\label{hyp3}
\end{equation}
Using $\Omega_{\cal N}$, the thermodynamic quantities can be obtained in the
standard way.

The leptons are included in the model as noninteracting particles,
since their interactions give negligible contributions compared to
those of their free Fermi gas parts.

For the kaon sector, we use a Lagrangian that contains the usual
kinetic energy and mass terms along with the meson interactions 
\citep[paper II]{GS99}.
Kaons are coupled to the meson fields through  minimal coupling;
specifically,
\begin{eqnarray}
L_K &=& {\cal D}^*_\mu K^+ {\cal D}^\mu K^- - m_K^{*2} K^+ K^- \,,
\end{eqnarray}
where the vector fields are coupled via the standard form \be {\cal
D}_\mu = \partial_\mu + i g_{\omega K} \omega_\mu + i g_{\rho K}
\gamma^{\mu}{\bf{b}}_{\mu}\cdot{\bf t} \ee and $m_K^{*} = m_K -
\frac{1}{2} g_{\sigma K} \sigma$ is the effective kaon mass.

In the mean field approach, the thermodynamic potential per unit volume 
in the kaon sector is (paper II) 
\begin{equation}
\frac{\Omega_K}{V}=\thalf (f\theta)^2(m_K^{*2}-(\mu+X_0)^2)
+T\int\limits_0^{\infty}\frac{d^3p}{(2\pi)^3}\left[
\ln(1-e^{-\beta(\omega^--\mu)})+
\ln(1-e^{-\beta(\omega^++\mu)})\right]\;,\label{zkexch}
\end{equation}
where $ X_0 = g_{\omega K} \omega_0 + g_{\rho K}b_0$, 
the Bose occupation probability $f_B(x)=(e^{\beta x}-1)^{-1}$, $\omega^{\pm} = 
{\sqrt {p^2+m_K^{*^2}}} \pm X_0$, 
$f=93$ MeV is the pion decay constant and the condensate amplitude,
$\theta$, can be found by extremization of the partition function.
This yields the solution $\theta=0$ (no condensate) or, if a
condensate exists, the equation \be m_K^{*} = \mu_K + X_0
\label{cond}
\ee
where $\mu_K$ is the kaon chemical potential.  In beta-stable stellar
matter the conditions of charge neutrality
\be
\label{cons1}
\sum_B q_B n_B - n_e - n_K = 0
\ee
and chemical equilibrium
\begin{eqnarray}
\mu_i &=& b_i\mu_n - q_i(\mu_l-\mu_{\nu_\ell}) \\
\mu_K &=& \mu_n - \mu_p
\label{tbeta}
\end{eqnarray}
are also fulfilled.

The kaon condensate is assumed to appear by forming a mixed phase with
the baryons satisfying Gibbs' rules for phase equilibrium \citep{Gibbs}.
Matter in this mixed phase is in mechanical, thermal and
chemical equilibrium, so that \be p^I=p^{II} \,, \quad T^I=T^{II}\,,
\quad \mu_i^I=\mu_i^{II}\,, \ee where the superscripts I and II
denote the nucleon and kaon condensate phases, respectively.  The
conditions of global charge neutrality and baryon number conservation
are imposed through the relations \be \chi q^I + (1-\chi) q^{II} &=& 0
\nonumber \\ \chi n_B^I + (1-\chi) n_B^{II} &=& n_B \,, \ee where
$\chi$ denotes the volume fraction of nucleonic phase, $q$ the charge
density, and $n_B$ the baryon density.  We ignore the fact that the
phase with the smallest volume fraction forms finite-size droplets; 
in general, this would tend to decrease the extent of
the mixed phase region. Further general consequences of imposing
Gibbs' rules in a multicomponent system are that the pressure varies
continuously with density in the mixed phase and that the charge
densities must have opposite signs in the two phases to satisfy global
charge neutrality.  We note, however, that not all choices of
nucleon-nucleon and kaon-nucleon interactions permit the Gibbs' rules
to be satisfied (for an example of such an exception, see paper II).
The models chosen in this work {\it do} allow the Gibbs' rules to be
fulfilled at zero and finite temperatures and in the presence of
trapped neutrinos.

The nucleon-meson couplings are determined by adjusting them to
reproduce the properties of equilibrium nucleonic matter at $T=0$.  We
use the numerical values used by \citet{GM91},
i.e., equilibrium density $n_0=0.153$ fm$^{-3}$, equilibrium energy
per particle of symmetric nuclear $E/A=-16.3$ MeV, effective mass
$M^*=0.78M$, compression modulus $K_0=240$ MeV, and symmetry energy
$a_{sym}=32.5$ MeV. These values yield the coupling constants
$g_\sigma/m_\sigma = 3.1507~{\rm fm},~ g_\omega/m_\omega = 2.1954~{\rm
fm},~g_\rho/m_\rho = 2.1888,~b=0.008659$, and $c=-0.002421$.

The kaon-meson couplings $g_{\sigma K}$ and $g_{\omega K}$ are related
to the magnitude of the kaon optical potential $U_K$ at the saturation
density $n_0$ of isospin symmetric nuclear matter: \be U_K(n_0) = -
g_{\sigma K} {\sigma(n_0)} - g_{\omega K} \omega_0(n_0).  \ee Fits to
kaonic atom data have yielded values in the range $-(50-200)$ MeV
\citep{FGB94,Fri99,WW97,RO00,BGN00}. We use
$g_{\omega K} = g_{\omega N}/3$ and $g_{\rho K} = g_{\rho N}/2$ on the
basis of simple quark and isospin counting.  Given the uncertainty in
the magnitude of $|U_K|$, consequences for several values of $|U_K|$
were explored in paper II.  Moderate values of $|U_K|$ generally produce a
second order phase transition and, therefore, lead to moderate effects on
the gross properties of stellar structure.  Values  in excess of
100 MeV were found necessary for a first order phase transition to
occur; in this case kaon condensation occurs at a relatively low
density with an extended mixed phase region, which leads to more
pronounced effects on the structure due to a significant softening of
the EOS.

\section{COMPOSITION AND STRUCTURE OF PNS WITH KAON CONDENSATES}

In Figure \ref{kaon-pre} we show the variation of pressure with baryon
density for three different choices of entropy per baryon, $s$, and
lepton fraction $Y_L=(n_e+n_\mu+n_\nu)/n_b$. The optical potential
$U_K$ is set to $-120$ MeV.  In an evolving PNS, $s$ and $Y_L$ are not
constant, but these three choices are reasonable approximations to the
ambient conditions before ($s=1$, $Y_{L}=0.35$) and after ($s=2$,
$Y_{\nu}=0$) the deleptonization stage, as well as for a cold,
neutrino-free ($s=0$, $Y_{\nu}=0$) neutron star.   The dotted line
corresponds to lepton--rich matter, the dashed line corresponds to hot
neutrino-free matter and the solid line corresponds to cold
neutrino-free matter.  Results obtained from Gibbs' construction are
shown by thick lines while those for the pure phases are shown by thin
lines.

Applying Gibbs' rules eliminates the region of negative
compressibility (this is more evident in the case of cold matter), and
also decreases the critical density for the onset of kaon
condensation.  Above the threshold density for condensation, the kaon
concentration builds up rapidly, which results in a significant
softening of the EOS.  The effects of trapped neutrinos are similar to
those obtained for hyperons in \citet{Pra97} and paper I. The
most notable feature due to the presence of trapped neutrinos is that
the critical density for kaon condensation is much higher in the case
neutrinos are trapped than in the case of neutrinos having left the
star. Notice that the extent of a  Maxwell construction would also be 
largely reduced in neutrino-trapped matter, and the differences between 
both kinds of mixed phase, Maxwell and Gibbs, are very small.

The phase boundaries of the different phases are displayed in Figure
\ref{kaon-bdy} in a $Y_L$--$n_B$ plane for an optical potential $U_K$
of --100 MeV (left) and --120 MeV (right), respectively.  The
nucleonic phase, the pure kaon matter phase, and the mixed phase are
labelled I, II, and III, respectively. Solid lines mark the phase
transition at zero temperature and dashed lines mark the phase
transition at an entropy per baryon of $s=1$.  Note that finite
entropy effects are small and do not affect significantly the phase
transition density.  The dash-dotted line shows the electron fraction
$Y_e$ as a function of density in cold, catalyzed matter (for which
$Y_L=Y_e$), which is the final evolutionary state. The region to the
left of this line corresponds to negative neutrino chemical potentials
and cannot be reached during normal evolutions.  The solid and dashed
lines, which separate the pure phases from the mixed phase, vary
roughly linearly with the lepton fraction.  Also notice the large,
and nearly constant, densities of the boundary between the mixed phase III
and the pure kaon phase II.  These densities, for the cases shown, lie
above the central densities of the maximum mass stars, so that region
II does not generally exist in proto-neutron stars (see Figure
\ref{kaon-mr} below).  The effect of increasing the lepton number is
to reduce the size of the mixed phase (which in fact shrinks to become
a second order phase transition for $Y_L>0.4$ and $U_K=-100$ MeV) and
to shift the critical density to higher densities.  A similar effect
is produced by decreasing the magnitude of the optical potential.

Figure \ref{kaon-mus} shows the baryon (upper panel) and lepton (lower
panel) chemical potentials for the three ambient conditions employed
in Figure \ref{kaon-pre}.  In neutrino-free matter, we observe a rapid
decrease of the electron chemical potential produced by the gradual
substitution of electrons by kaons. When neutrinos are trapped,
$\beta$-equilibrium prevents the electron chemical potential from
decreasing, because the neutrino chemical potential increases with
density.  In Figure \ref{kaon-comp} the compositions are for the same
ambient conditions of $Y_L$ and $s$ as in Figure \ref{kaon-pre}.  A
remarkable effect is the very high proton fraction reached at high
densities, higher than the neutron fraction at densities above 0.8
fm$^{-3}$. The figure also displays how both thermal and
lepton-trapping effects displace the critical density to higher
values, the effects of having neutrinos in the system being more
important than those of finite temperature.

In Figure \ref{kaon-mr} we show the baryon mass ($M_B$, upper panels)
and gravitational mass ($M_G$, lower panels) versus central baryon
density, for $U_K=-120$ MeV (left panels) and --100 MeV (right
panels), respectively.  The onset of kaon condensation is shown by the
diamond on each curve.  In each case, the maximum mass of the cold
neutrino-free star is below the maximum mass of the hot star,
irrespective of whether it has neutrino trapping. This figure
indicates the existence of a mass window for metastability
($1.7<M_B<2.0$ M$_\odot$ for $U_K=-120$ MeV and $2.0<M_B<2.2 {\rm
M}_\odot$ for $U_K=-100$ MeV) in which a PNS will become unstable at
the end of its Kelvin-Helmholtz stage.  A similar effect was found in
paper I due to the presence of hyperons in matter.  It should be
emphasized that this figure uses constant $s$ and $Y_L$ profiles
within the star.  In fact, these quantities are not constant in the
interior, so the precise mass window depends upon the initial profile
and the evolution.  We have found, however, that the mass window
exhibited in this figure is a reasonable approximation.

The main effect of lowering $|U_K|$ is a reduction in the size of the
metastability window and its displacement to higher values, and this
is shown in Figure \ref{mb-uk}. The value of $U_K$ must be limited to
values greater than about --126 MeV in order that the binary pulsar
mass constraint, 1.44 M$_\odot$, not be violated. For values of $U_K$
greater than about --80 MeV, it is also clear that metastable models
cannot exist.  Another interesting feature is that the maximum masses
of hot, neutrino-free models are nearly the same (for $U_K=-120$ MeV)
or even larger (for $U_K=-100$ MeV) than those of lepton-rich
configurations, preventing metastable stars from collapsing upon
deleptonization and delaying the onset of instability until the end of
the cooling stage.  This means that kaon-induced instabilities cannot
occur very quickly, as can be the case with hyperon induced collapses.

\section {NEUTRINO OPACITIES}

For neutral and charged current processes involving nucleons and
leptons, we utilize the recent advances made in the works of 
\citet{RPL98,Red99,BS98,BS99}. In these works, neutrino cross
sections for arbitrary degeneracies were calculated, including the
effects of interactions and collective excitations in beta stable
isospin asymmetric matter.  These opacities form the basis of our
baseline calculations against which modifications to the opacities due
to the presence of a kaon condensate are assessed. 

In bulk equilibrium and in
the degenerate limit, the factors by which the normal reaction rates
must be modified in the presence of a charged kaon condensate are
\citep{Tho95}: 
\be 
\nu_e + n(K) &\rightarrow & n(K) + e^-~: ~  \frac 14 \sin^2\theta
\tan^2\theta_C  \nonumber \\ 
\nu_e + p(K) &\rightarrow &p(K) + e^-~: ~ \sin^2\theta \tan^2\theta_C 
\nonumber \\ 
\nu_e + n(K) &\rightarrow & p(K) +e^-~: ~ \cos^2(\theta/2) 
\ee 
Here,
$n(K)$ denotes an excitation which is a superposition of a neutron and
a proton, and reduces to a free neutron or proton in the absence of a
charged kaon (or pion) condensate, $\theta$ denotes the amplitude of
the condensate and $\theta_C$ is the Cabibbo angle ($\sin \theta_c =0.23$).
The factors in this equation are all 
less than unity.  The opacity for the neutral
current scattering processes involving nucleons in the presence of a
kaon condensate has also been investigated recently by \citet{JM98}
and by \citet{Red98}, who included the SU(3) singlet
contribution.  The general result is that a significant reduction in
the cross section is caused by the presence of a kaon condensate
formed in a bulk medium.

However, the situation is radically different for the opacity due to a
kaon condensation in a mixed phase, in which finite-size effects may
be important \citep{GS98,RBP00,CGS00}.
\citet{RBP00} found that coherent scattering of neutrinos
off kaon-condensed droplets in the mixed phase increases the neutrino
cross section, relative to bulk nucleonic matter, by factors of
10--20.  The overall enhancement, however, is sensitive to the droplet
size, which depends on the surface tension.

We consider two extreme models: one in which the cross section or
opacity is significantly smaller (20 times) than that of the
neutrino-nucleon processes to account for a condensate in bulk matter,
and one in which the cross section is significantly larger (20 times),
to account for possible effects of a droplet phase.  We apply these
factors to the overall opacity, while in fact the reduction factor for
the bulk condensate should apply only to the volume occupied by the
kaon-condensed phase.  The mixed phase occupies only a small fraction ($\sim
0.2$) of the total volume even in the most massive PNSs, which
mitigates considerably the large reductions in the opacity in this
case. We wanted to ensure that our models will unmistakably bracket
the true situation.  Our results conclusively demonstrate that even large
modifications to the opacity caused by kaon condensation do not have a
pronounced effect on neutrino signals.

\section{EVOLUTION}

We now proceed to the evolution of a PNS in which a kaon condensate
forms.  We will focus on the case in which $U_K=-120$ MeV, for which
the effects of kaon condensation are relatively pronounced.
Calculations \citep{Pra96,MTI00} of
the growth rate for a kaon condensate indicate a timescale of less or
about $10^{-4}$ s at the temperatures of interest (above a few
MeV). Since this is much shorter than the Kelvin-Helmholtz time-scale
(several s), we will assume matter to reach chemical equilibrium
instantaneously.  However, for hydrodynamical studies, such as those
involving collapse to a black hole or pulsations due to core quakes,
the kaon condensation formation timescale might have to be considered.

Initial entropy per baryon $s$ and lepton fraction $Y_L$ profiles are
taken to be the same as in paper I, where we studied the evolution of
PNSs containing either pure nucleons or nucleon/hyperon mixtures. The
numerical code used to perform the simulations was also described in
paper I; the code is based on a Henyey-like scheme in which the
structure and transport equations are solved in sequential steps and
then corrector steps are taken until convergence is reached.

For $M_B<1.6$ M$_\odot$, the central density does not exceed the
critical value for kaon condensation (for the set of parameters
employed), and the evolution is identical to that of pure nucleon
($np$) models described in paper I. For stars with $1.6 < M_B/{\rm
M}_\odot<2.05$, a kaon condensate will form during the evolution. We
can classify the PNSs with a kaon condensate core in two subclasses:
i) stable, those with $M_B<1.75$ M$_\odot$, which is the maximum mass
of cold, neutrino-free neutron stars (see Figure \ref{kaon-mr}); ii)
metastable, those configurations with $1.75<M_B/{\rm M}_\odot<2.05$.

The evolution of stars containing kaon condensates ($npK$) 
deviates from those of $np$ stars when the mixed phase with a condensate 
forms at the core.  The size of this mixed phase grows on a diffusion
time scale, i.e., the average time it takes for the neutrinos produced
during kaonization to escape the inner core.  In Figure
\ref{kaon-struc}, we show profiles of particle fractions as a function
of radius for a stable star with $M_B=1.7 {\rm
M}_\odot$ towards the end of the Kelvin-Helmholtz stage ($t=60$ s).
At this time, the star is at
the beginning of its long-term cooling phase 
during which its final catalyzed state is achieved.  Notice that the size of
the core is about 3 km, but it contains only a small fraction
($\approx 0.05 {\rm M}_\odot$) of the star's total mass.

In Figure \ref{evol1}, the evolution of stable (1.7 M$_\odot$) and
metastable stars are compared.  Both the central baryon number
densities and kaon fractions $Y_K$ are displayed. In each case, the
time at which kaon condensation occurs is indicated by a diamond.
Asterisks mark the times at which the evolution of metastable stars
could not be further followed in our simulations, i.e., when a
configuration in hydrostatic equilibrium could not be found.  At this
time, the PNS is unstable to gravitational collapse into a black hole.
For the stable star, kaons appear after about 40 s.  Thereafter, the
star's central density increases in a short interval, about 5 s, until
a new stationary state with a mixed phase is reached.  The evolution
of the metastable stars is qualitatively different, inasmuch as the
central density increases monotonically from the time the condensate
appears to the time of gravitational collapse.

It is interesting that the lifetimes in all cases shown lie
in the narrow range 40--70 s (see Fig. \ref{evol1}).  They decrease
mildly with increasing $M_B$.  Further insight is obtained by
examining Figures \ref{evol2} and \ref{evol3}. In Figure \ref{evol2}
the evolution of the central baryon density and kaon fraction are
shown as functions of the central value of $Y_L$.  The initial
contraction of the star occurs at nearly constant lepton fraction, as
can be seen from the vertical trajectories of the central
density. This is followed by a stage in which the lepton fraction
decreases while little variation in the central density occurs, until
kaon condensation takes place (marked by diamonds in the figures).
Thereafter, both the central density and the kaon fraction increase
until the PNS reaches its final stage, either the cold-deleptonized
configuration or the unstable configuration that collapses to a black
hole.  

Although the densities and lepton fractions for which condensation
occurs are different for stars with different masses, the typical time
scales are similar, as is evident from Figure \ref{evol3} in which the
evolution of the electron and neutrino concentrations, $Y_e$ (top
panel), $Y_{\nu}$ (middle panel), respectively, and the temperature
(bottom panel) at the star's center are shown. For all cases, the
lepton concentrations exhibit a {\em plateau stage} until about 30-40
s, and remain high enough to prevent the formation of a kaon
condensate.  Once the condensate is formed, $Y_\nu$ may increase as
kaons gradually replace electrons in the inner core.

Notice that the {plateau stage} is also present in the evolution of
stars containing only nucleons, as was shown in paper I.  Until a kaon
condensate appears, the evolution of stars considered in this work
proceeds similarly to that of a nucleons-only star because the number
of thermal kaons present is rather small.  The plateau is caused by an
increase in neutrino opacity due to high temperatures, which reduces
the leakage of neutrinos from the core, in turn maintaining a high
lepton fraction.  Consequently, kaonization of matter is delayed until
the end of this epoch, due to the dependence of the threshold density
on $Y_L$.  This explains why kaon-condensed stars of different masses
become unstable at approximately the same time.  This is qualitatively
different from metastable hyperon-rich stars studied in paper I, for
which the lifetime is a much more sensitive function of mass.  This is
displayed in Figure \ref{ttc}, in which the lifetimes as a function of
$M_B$ for stars containing hyperons ($npH$) and $npK$ stars are
compared.  In both cases, the larger the mass, the shorter the
lifetime.  For kaon-rich PNSs, however, the collapse is delayed until
the final stage of the Kelvin-Helmholtz epoch, while this is not
necessarily the case for hyperon-rich stars.

Strangeness appearing in the form of a mixed phase of strange quark
matter also leads to metastability.  Although quark matter is also
suppressed by trapped neutrinos \citep{PCL95,SPL00},
the transition to quark matter can occur at lower densities than 
the most optimistic kaon case, and the dependence of
the threshold density on $Y_L$ is less steep than that for kaons.  Thus, it
is an expectation that metastability due to the appearance of quarks,
as for the case of hyperons, might be able to occur relatively
quickly.  Calculations of PNS evolution with a mixed phase of quark
matter, including the possible effects of quark matter superfluidity
\citep{CR00} are currently in progress and will be reported
separately.

An interesting question concerns how short the lifetime of a
kaon-condensed metastable star can be. For the case $U_K=-120$ MeV,
this time is constrained by the fact that metastability disappears for
$M_B\approx 2.05 {\rm M}_\odot$. For smaller magnitudes of the optical
potential, the lifetimes are larger since the critical densities are
larger; these densities are reached only after longer times.
Conversely, increasing the magnitude of $U_K$ decreases the lifetime.
The magnitude of the optical potential is limited, however, by the
binary pulsar mass constraint that the maximum mass must exceed 1.44
${\rm M}_\odot$, or $|U_K|<126$ MeV for our interaction.  In this
case, the lifetime is found to be about 40 s.  This leads to the
interesting conclusion that should metastability from a PNS with a
timescale of, say, 30 s or less be observed, it would be
evidence for the existence of hyperonic or quark matter rather than 
for a kaon condensate.

\section{LUMINOSITIES AND SIGNALS IN TERRESTRIAL DETECTORS}

The time dependence of the neutrino signal from a supernova is the
only observable means through which the physical processes occurring
in the high density inner core of a PNS may be inferred.  Our study is
more applicable to times longer than approximately 1 s after core bounce,
after which effects of dynamics and accretion become unimportant.
Studies of the neutrino signal during the first second, during which
approximately 1/3 of the neutrino energy is emitted, require
more accurate  techniques for neutrino
transport \citep{MMBG98,YJS99,BYPET00} coupled with recent developments
in supernova simulations \citep{Mez00,RJ00}.

For simplicity, we limit
our study to the signal produced by electron antineutrinos absorbed
onto protons in a pure water detector, for which the cross section is
$\sigma(E_\nu)=\sigma_0E_\nu^2$ where $E_\nu$ is the $\bar\nu_e$
energy in MeV and $\sigma_0=9.3\times 10^{-44}$ cm$^2$.  This was the
major reaction observed during the SN 1987A neutrino burst and it is
by far the process with the highest rate in the newer detectors.
Assuming that the neutrinos leaving the PNS have a Fermi-Dirac
spectrum with zero chemical potential, knowledge of the neutrino
luminosities and average temperatures permit the signal from a
supernova to be calculated according to (paper I)
\begin{equation}
{dN\over dt}={\sigma_0n_p\over 4\pi D^2}{\cal M} {G_4(E_{th},
T_\nu^{\infty})\over F_3(0)} T_\nu^{\infty} L_{\bar\nu}  \,. 
\label{rate}
\end{equation}
In this equation, $L_{\bar\nu}$ is the antineutrino luminosity
(assumed to be 1/6 of the total neutrino luminosity), $n_p=6.7\times
10^{28}$ free protons per kiloton of water, $D$ is the distance to
the supernova, ${\cal M}$ is the fiducial mass of the detector, and
$T_\nu^{\infty}$ is the redshifted neutrino temperature.
Also, $F_3(0)=7\pi^4/120$ is an ordinary Fermi integral and $G_i(E_{th},T)$
denotes a modified, truncated Fermi integral:
\begin{equation}
G_i(E_{th},T)=\int_{E_{th}}^\infty dz\, z^i W(zT)(1+e^z)^{-1} \,
\end{equation}
with $E_{th}$ being the detector threshold and $W(E)$ the detector
efficiency.  We studied the signal observed in KII and IMB, whose
masses, thresholds and efficiencies are taken to be the same as in
\citet{LY89}, except that 6 ktons is used for the IMB mass, as 
discussed by \citep{SB90}.  We have assumed thresholds and fiducial 
masses of 5.6 (2) MeV and 22.4 (1.7) ktons for the SuperK (SNO) detectors,
respectively, and also assumed a detector efficiency of unity above
the threshold for both.  Although we assumed that PNS neutrinos have a
zero chemical potential spectrum, the formula (\ref{rate}) does not
depend very sensitively on this assumption \citet{LY89}.

The luminosity $L_{BG}$ corresponding to a given background rate below
which the signal is lost, $(dN/dt)_{BG}$, can be obtained from
inverting equation (\ref{rate}):
\begin{equation}
L_{BG} = 105.0 {\left( dN\over dt \right)}_{BG} 
\frac{ {(D/10{\rm~kpc})}^2 }{ ({\cal M}/{\rm kton}) } F(T_{\bar\nu}^{\infty})
\,,
\end{equation}
where $F(T_{\bar\nu}^\infty)$ also depends on the threshold and
efficiency of each detector. We have obtained the following analytical
fits \be F(T) = \left\{ {\matrix{ \displaystyle{
\frac{1+\exp{[2.0(1.2-T)]}}{25.5(T-0.75)} } & {\rm KII},~ T>0.75 \cr &
\cr (-5.27 T + 2.57 T^2)^{-1} & {\rm IMB}, ~T>2.05 \cr & \cr
\displaystyle{ \frac{1+\exp{[2.25(1.2-T)]}}{23.3 T} } & {\rm SuperK}
\cr & \cr \displaystyle{ \frac{1+\exp{[6.0(0.4-T)]}}{23.3 T} } & {\rm
SNO} \cr }} \right.  \ee where $T$ is in MeV.  In general, we will
assume that the limiting background rate is about 0.2 Hz (one count
every 5 s), since the corresponding time between counts is then about
10\% of the entire signal time.  A better estimate would require an
in-depth statistical analysis that is beyond the scope of this paper.

In Figure \ref{lum1} the evolution of the total neutrino energy
luminosity is shown for different models. Notice that the drop in the
luminosity for the stable star (solid line), associated with the end
of the Kelvin-Helmholtz epoch, occurs at approximately the same time as
for the metastable stars with somewhat higher masses.  In all cases,
the total luminosity at the end of the simulations is below $10^{51}$
erg/s.  The two upper shaded bands correspond to SN 1987A detection
limits with KII and IMB, and the lower bands correspond to detection
limits in SNO and SuperK for a future galactic supernova at a distance
of 8.5 kpc.  The width of the bands represents the uncertainty in the
$\bar\nu_e$ average energy because of the limitations of the
flux-limited diffusion approximation, as discussed in paper I.  The
times when these limits intersect the model luminosities indicate the
approximate times at which the count rate drops below the background
rate $(dN/dt)_{BG}=0.2$ Hz.

The poor statistics in the case of SN 1987A precluded a precise
estimate of the PNS mass.  Nevertheless, had a collapse to a black
hole occurred in this case, it must have happened after the detection
of neutrinos ended. Thus the SN 1987A signal is compatible with a late
kaonization-induced collapse, as well as a collapse due to
hyperonization or to the formation of a quark core.  More information
would be extracted from the detection of a galactic SN with the new
generation of neutrino detectors.

In SNO, about 400 counts are expected for electron antineutrinos from
a supernova located at 8.5 kpc.  The statistics would therefore be
improved significantly compared to the observations of SN 1987A.  A
sufficiently massive PNS with a kaon condensate becomes metastable,
and the neutrino signal terminates, before the signal decreases below
the assumed background.  In SuperK,
however, up to 6000 events are expected for the same conditions
(because of the larger fiducial mass) and the effects of metastability
due to condensate formation in lower mass stars would be observable.

It is interesting to assess the role of neutrino opacities in the
presence of kaon condensates.  For this purpose, we calculated the
luminosities with two extreme assumptions concerning the opacities,
namely, the opacity is (i) 20 times smaller than the baseline opacity,
corresponding to the bulk case \citep{Tho95,JM98},
and (ii) 20 times larger than the baseline opacity, corresponding to
finite-size effects in the mixed phase \citep{RBP00}.

In Fig {\ref{opac}}, we show the temporal evolution of the central
baryon density (top panel) and the total luminosity (bottom panel),
for models with and without the correction to the opacity. In
addition, we also show results corresponding to the case of pure $np$
matter, in which the kaon condensate is not allowed.  The magnitude of
the opacity in the mixed phase chiefly controls the growth rate of the
central density (top panel), and hence of the condensate, with larger
opacities causing longer delays.  The effect of reduced opacities on
the luminosities (bottom panel) is barely distinguishable from the
baseline case.  In the case of enhanced opacities, for the stable
star, the luminosity is first reduced by about 10-20 \% for 10 s
following the appearance of kaons at $t\sim 45$ s, but eventually
exceeds that of the baseline case.  Nevertheless there is only a small
change in the lifetime produced by even large opacity changes.  The
metastable star in Figure \ref{opac} with 2M$_\odot$ collapses before
much information about the physics in the core reaches the
neutrinosphere.  It is also remarkable that the largest differences in
the luminosities between the models with and without a condensate, but
for the same opacity, are smaller than about 2\%.

It is important to note, however, that by the time significant effects
due to a kaon condensate become visible ($t>50$ s), our treatment of
neutrino transport becomes suspect since neutrinos in the
interior are entering the semi-transparent regime. This is reflected
in the location of the neutrinosphere, which begins to fall well below
the surface of the star for these times, and in the total optical
depth of the star, which is no longer large.  These affect our
estimates of both the neutrino signal and the effective detector
background.

The expected neutrino signals in the SuperK detector from a supernova
at $8.5$ kpc are compared in Fig. \ref{hypc} for a sample of stable
PNSs with different compositions and masses.  To show the effects of
compositional changes, stars with $M_B=1.7$ M$_\odot$ for $np$, $npH$
and $npK$ cases are shown, and to show the effects of small changes
in the stellar mass, an $M_B=1.8$ M$_\odot$ for the case of $npK$ is
also displayed. The upper panel shows the count rate and the lower
panel shows the integrated number of counts.  It is clear that, for
the same stellar mass and for stable models, the effect of having a
kaon condensate core is smaller than the effect of hyperons: the $np$
case is indistinguishable from the $npK$ case.  A change in the
assumed mass of the PNSs of about 0.1 M$_\odot$ has about twice the
effect on the number of counts as the change produced by introducing
hyperons.  Therefore, unless an independent method of identifying the
PNS mass is available, small differences in the assumed PNS mass can
mask compositional effects.  

Furthermore, although a clear signature of the presence of strangeness
in PNSs occurs due to the collapse of metastable models, there is no
unique feature that establishes the presence of a kaon condensate.  If
the lifetime of the metastable star is less than about 30 s, however,
it appears unlikely that the instability is due to kaon condensation.

The main conclusions arising from these comparisons is that the
presence of a kaon condensate would be difficult to establish in an
unambiguous fashion.  For stars that are stable, the small modulations
in the time dependence of the neutrino luminosity or count rate due to
kaon condensation can be easily masked by differences in composition
(i.e., the presence of hyperons) or a small difference in the assumed
PNS mass compared to a nucleons-only case.  For stars that are
metastable, the sudden cessation in the neutrino signal is similar for
instabilities whether caused by kaons, hyperons or quarks.

\section{SUMMARY AND CONCLUSIONS}

We have studied the effect of kaon condensation on PNS evolution, by
employing an EOS which includes the effects of finite temperature and
neutrino trapping.  The phase transition from pure nucleonic matter to
the kaon condensate phase is described by means of Gibbs' rules for
phase equilibrium which permit a mixed phase. In the models explored
here, the central densities are not large enough to allow a pure
condensed phase to exist.

We can classify stars of different masses in three main groups: i)
stars in which the central density does not exceed the critical value
for kaon condensation, ii) stars that can form a mixed phase core at
the end of the Kelvin-Helmholtz epoch but remain stable, and, iii)
kaon condensed metastable stars that become unstable and collapse to a
black hole at the end of the Kelvin-Helmholtz epoch.  For stars in
which the effects of kaon condensation are small (this could be either
because the star's mass is low enough to permit only a very small
region of the mixed phase or because condensation occurs as a weak
second order phase transition), the differences of the predicted
neutrino signal compared to PNSs composed of pure nucleonic matter
are very small.

Our calculations show that the variations in the neutrino light curves
caused by the appearance of a kaon condensate in a stable star are
small, and are apparently insensitive to large variations in the
opacities assumed for them.  Relative to a star containing only
nucleons, the expected signal differs by an amount that is easily
masked by an assumed PNS mass difference of $0.01-0.02$ M$_\odot$.
This is in spite of the fact that, in some cases, a first order phase
transition appears at the star's center.  The manifestations of this
phase transition are minimized because of the long neutrino diffusion
times in the star's core and the Gibbs' character of the transition.
Both act in tandem to prevent either a ``core-quake'' or a secondary
neutrino burst from occurring during the Kelvin-Helmholtz epoch.

Observable signals of kaon condensation occur only in the case of
metastable stars that collapse to a black hole.  In this case, the
neutrino signal for a star closer than about 10 kpc is expected to
suddenly stop at a level well above that of the background in a
sufficiently massive detector with a low energy threshold such as
SuperK.  This is in contrast to the signal for a normal star of
similar mass for which the signal continues to fall until it is
obscured by the background.  The lifetime of kaon-condensed metastable
stars has a relatively small range, of order 50--70 s for the models
studied here, which is in sharp contrast to the case of hyperon-rich
metastable stars for which a significantly larger variation in the
lifetime (a few to over 100 s) was found.  This feature of kaon
condensation suggests that stars that destabilize rapidly cannot do so
because of kaons.

We determined the minimum lifetime for metastable stars with kaons to
be about 40 s, by examining the most favorable case for kaon
condensation, which is obtained by maximizing the magnitude of the
optical potential.  The maximum optical potential is limited by the
binary pulsar mass constraint, which limits the star's maximum
gravitational mass to a minimum value of 1.44 M$_\odot$.  Therefore,
should the neutrino signal from a future supernova abruptly terminate 
sooner than 40 s after the birth of
the PNS, it would be more consistent with a hyperon- or quark-induced
instability than one due to kaon condensation.

It is important to note that the collapse to a black hole in the case
of kaon condensation is delayed until the final stages of the
Kelvin-Helmholtz epoch, due to the large neutrino diffusion time in
the inner core.  Consequently, to distinguish between stable and
metastable kaon-rich stars through observations of a cessation of a
neutrino signal from a galactic supernovae is only possible using
sufficiently massive neutrino detectors with low energy thresholds and
low backgrounds, such as the current SNO and SuperK, and future
planned detectors.

\acknowledgements
This work was supported in part by the Spanish DGCYT grant
PB97-1432 and the U.S. Department of Energy under contract numbers
DOE/DE-FG02-87ER-40317 (JAP and JML) and  DOE/DE-FG02-88ER-40388 (MP).
JAP thanks J.M. Ib\'a\~nez for useful discussions and
encouragement.


{}

\newpage

\figcaption{\label{kaon-pre}}
{Total pressure as a function of baryon density $n_B$,
for different lepton fractions and entropy per baryon in matter with
a kaon condensate. Thin lines show results from the pure phases
while thick lines are results obtained employing  Gibbs' phase rules.}

\figcaption{\label{kaon-bdy}} {Phase boundaries between pure nucleonic
matter (I), pure kaon condensed matter (II) and a mixed phase (III) in
the $Y_L$--$n_B$ plane for $U_K=-100$ MeV (left panel) and $U_K=-120$
MeV (right panel). The solid line corresponds to $s=0$ and the dashed
line to $s=1$. The dashed-dotted line shows the baryon density as a
function of the lepton fraction for $s=0$, neutrino-free ($Y_L=Y_e$)
matter.}

\figcaption{\label{kaon-mus}}
{Baryon (top panel) and lepton (bottom panel) chemical potentials 
for the physical conditions shown in Figure \ref{kaon-pre}.}

\figcaption{\label{kaon-comp}} {Particle concentrations as a function
of baryon density $n_B$ for the physical conditions shown in Figure
\ref{kaon-pre}.}

\figcaption{\label{kaon-mr}} {Baryon (top) and gravitational (bottom)
mass versus central baryon density for $U_K=-120$ MeV (left) and --100
MeV (right), respectively, for the physical conditions shown in Figure
\ref{kaon-pre}.  Diamonds indicate the density above which kaon
condensation occurs.}

\figcaption{\label{mb-uk}} {The maximum baryon mass for the physical
conditions shown in Figure \ref{kaon-pre} as a function of the optical
potential $U_K$. Metastable stars cannot exist when the cold,
catalyzed maximum mass exceeds that of neutrino trapped
matter. The triangle indicates the binary pulsar gravitational mass
constraint $M_G^{max}=1.44$ M$_\odot$.}

\figcaption{\label{kaon-struc}} {Particle concentrations as a function
of the radius of a PNS containing a kaon condensate core.  The figure
is a snapshot at $t=60$ s, near the end of the Kelvin-Helmholtz stage.
Notice the increase of proton concentration in the presence of kaons.}

\figcaption{\label{evol1}} {Evolution of the central baryon density
$n_B$ and the kaon/baryon fraction $Y_K$ for stars with different
baryonic masses.  Solid lines correspond to a stable star with
$M_B=1.7 {\rm M}_\odot$; stars with larger masses are
metastable. Diamonds indicate when kaon condensation occurs at the
stellar center, and asterisks denote when metastable stars become
gravitationally unstable.}

\figcaption{\label{evol2}} {Evolutionary tracks of the central
densities and kaon fractions as functions of $Y_L$ for stars of
various baryon masses.  Diamonds and asterisks have the same meaning
as in Figure \ref{evol1}.}

\figcaption{\label{evol3}} {Evolution of the central temperature
(bottom) and neutrino (middle) and electron (top) fractions for
stars of various baryon masses.  Diamonds and asterisks have the same meaning
as in Figure \ref{evol1}.}

\figcaption{\label{ttc}} {Lifetimes of metastable stars as a function
of the stellar baryon mass.  Solid lines show results for PNSs
containing kaon-condensates and dashed lines show the results of paper
I for PNSs containing hyperons.}

\figcaption{\label{lum1}} {The evolution of the total neutrino
luminosity for stars of various baryon masses.  Line and symbol
designations are the same as in Figure \ref{evol1}.  Shaded bands
illustrate the limiting luminosities corresponding to a count rate of
0.2 Hz in all detectors, assuming a supernova distance of 50 kpc for
IMB and Kamioka, and 8.5 kpc for SNO and SuperK. The width of the
shaded regions represents uncertainties in the average neutrino energy
from the use of a diffusion scheme for neutrino transport.}

\figcaption{\label{opac}} {The evolution of the central baryon density
(top) and total neutrino luminosity (bottom) for two stars with
$M_B=1.7$ and 2.0 M$_\odot$, for different assumptions concerning
composition and opacity.  Solid lines refer to the baseline model of a
kaon-condensed star which assumes opacities corresponding to pure
nucleonic matter.  Dotted lines refer to stars containing only
nucleons.  Dashed (dot-dashed) lines refer to models of kaon-condensed
stars in which the opacity is increased (decreased) by a factor 20.
Asterisks indicate when metastable models become unstable to
gravitational collapse.  The shaded bands are as in Figure
\ref{lum1}.}

\figcaption{\label{hypc}} {The expected neutrino signal in the SuperK
detector from a supernova at $8.5$ kpc for PNSs with different
compositions and masses.  For the case $M_B=1.7$ M$_\odot$, $np$,
$npK$ and $npH$ compositions are compared.  For the $npK$ composition,
stars with two masses of (1.7 and 1.8 M$_\odot$) are compared. The
upper panel shows the count rate and the lower panel shows the
integrated number of counts. }

\newpage

\begin{figure}
\plotone{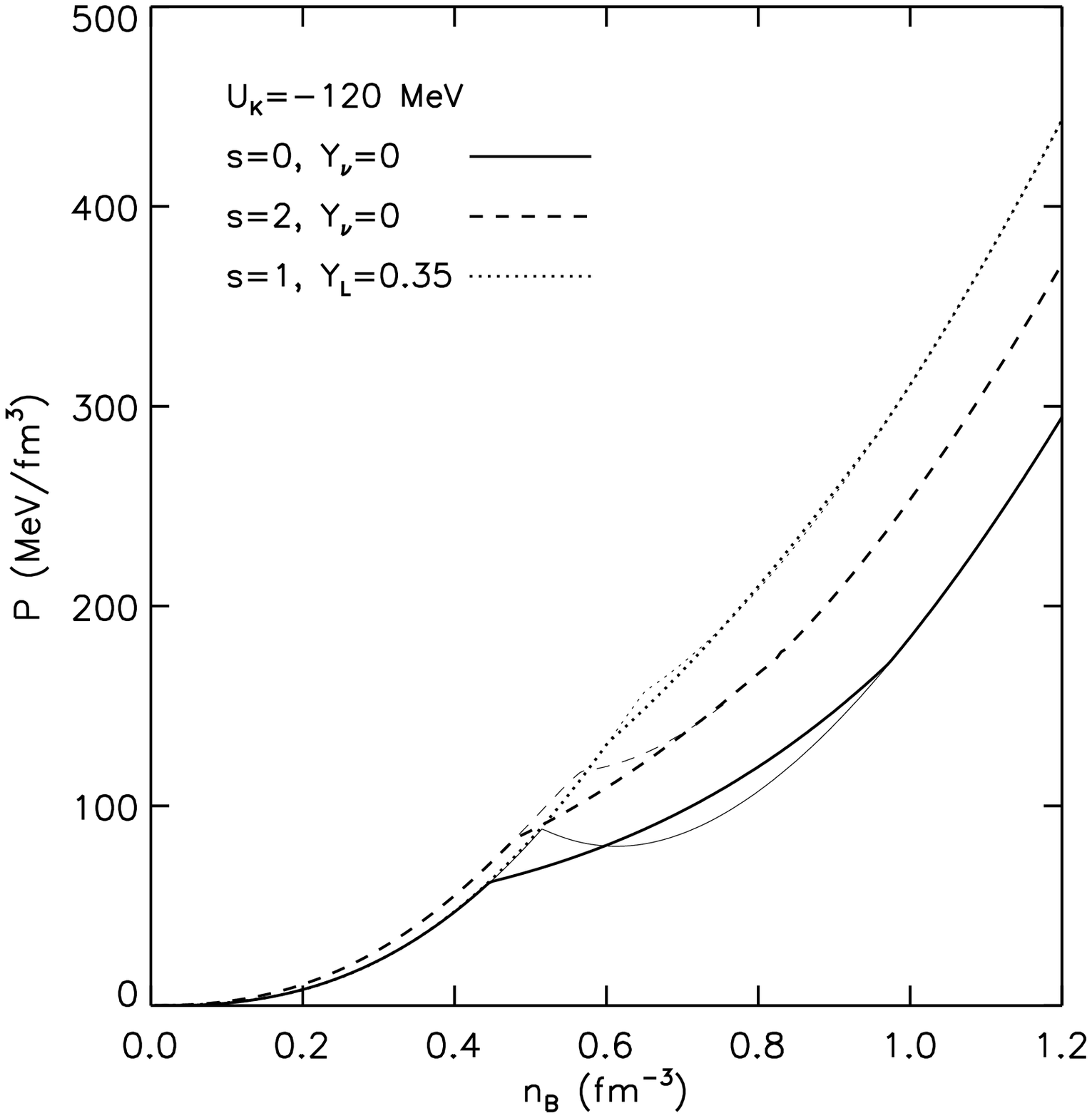}
\end{figure}

\newpage

\begin{figure}
\plotone{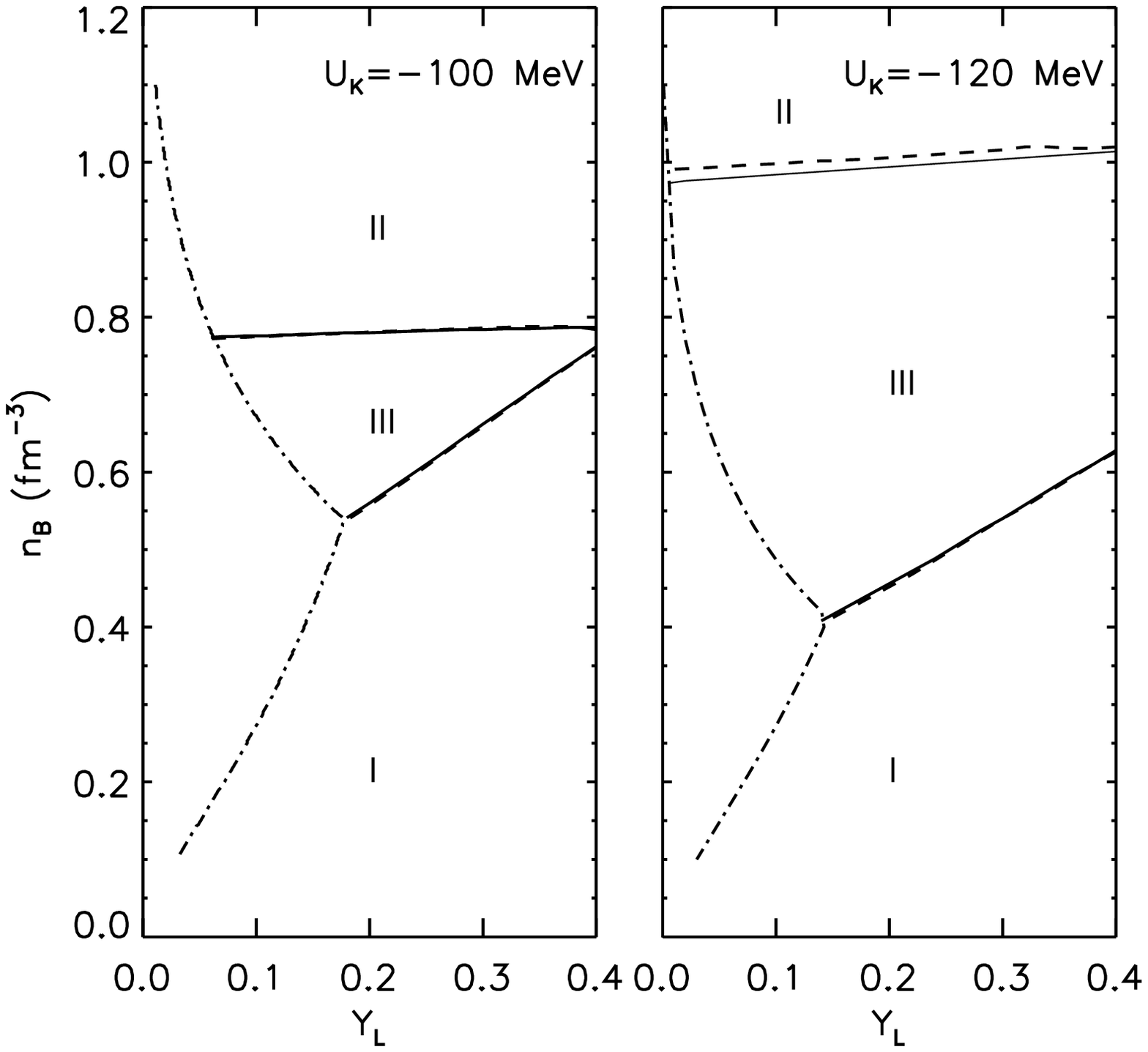}
\end{figure}

\newpage

\begin{figure}
\plotone{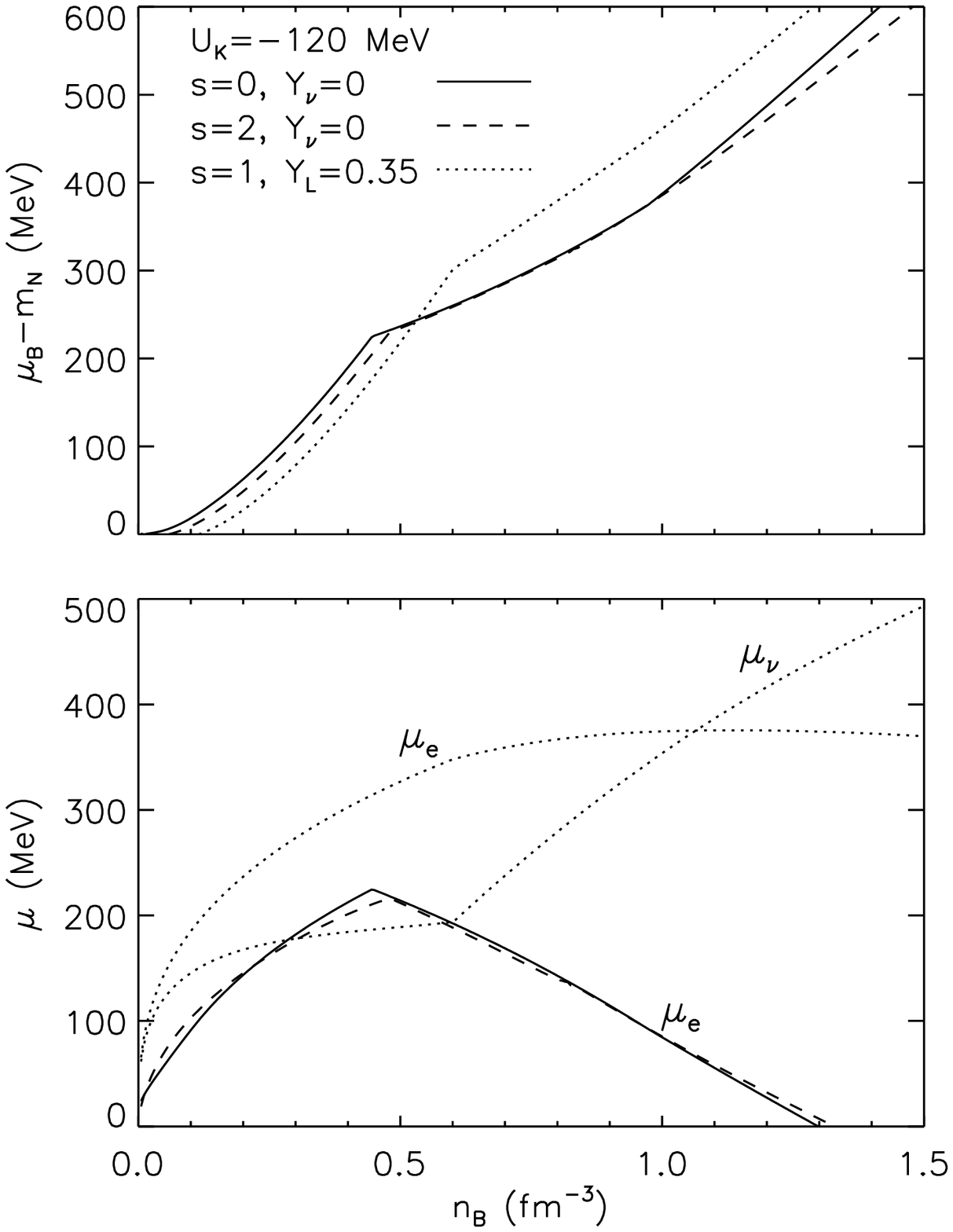}
\end{figure}

\newpage

\begin{figure}
\plotone{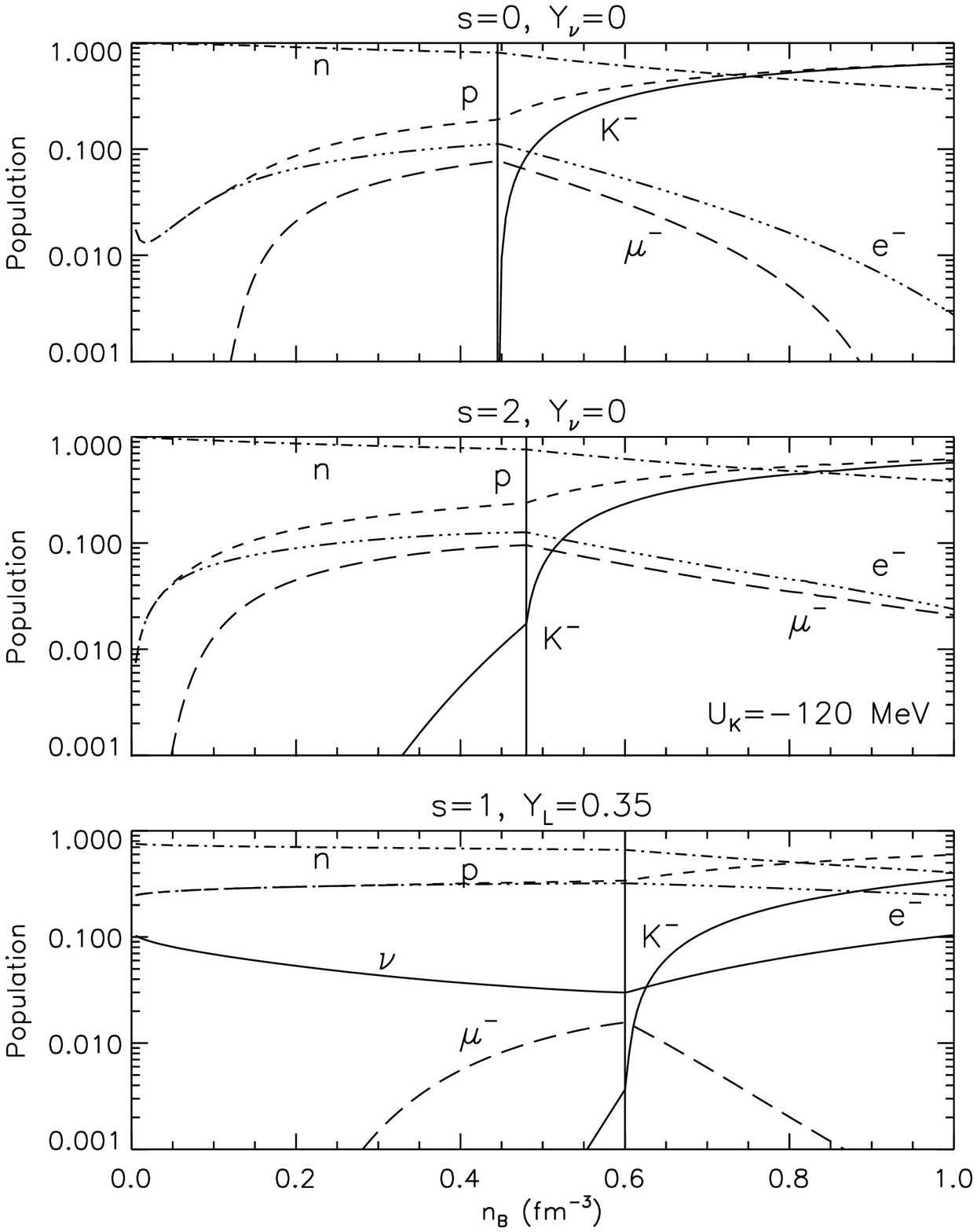}
\end{figure}

\newpage

\begin{figure}
\plotone{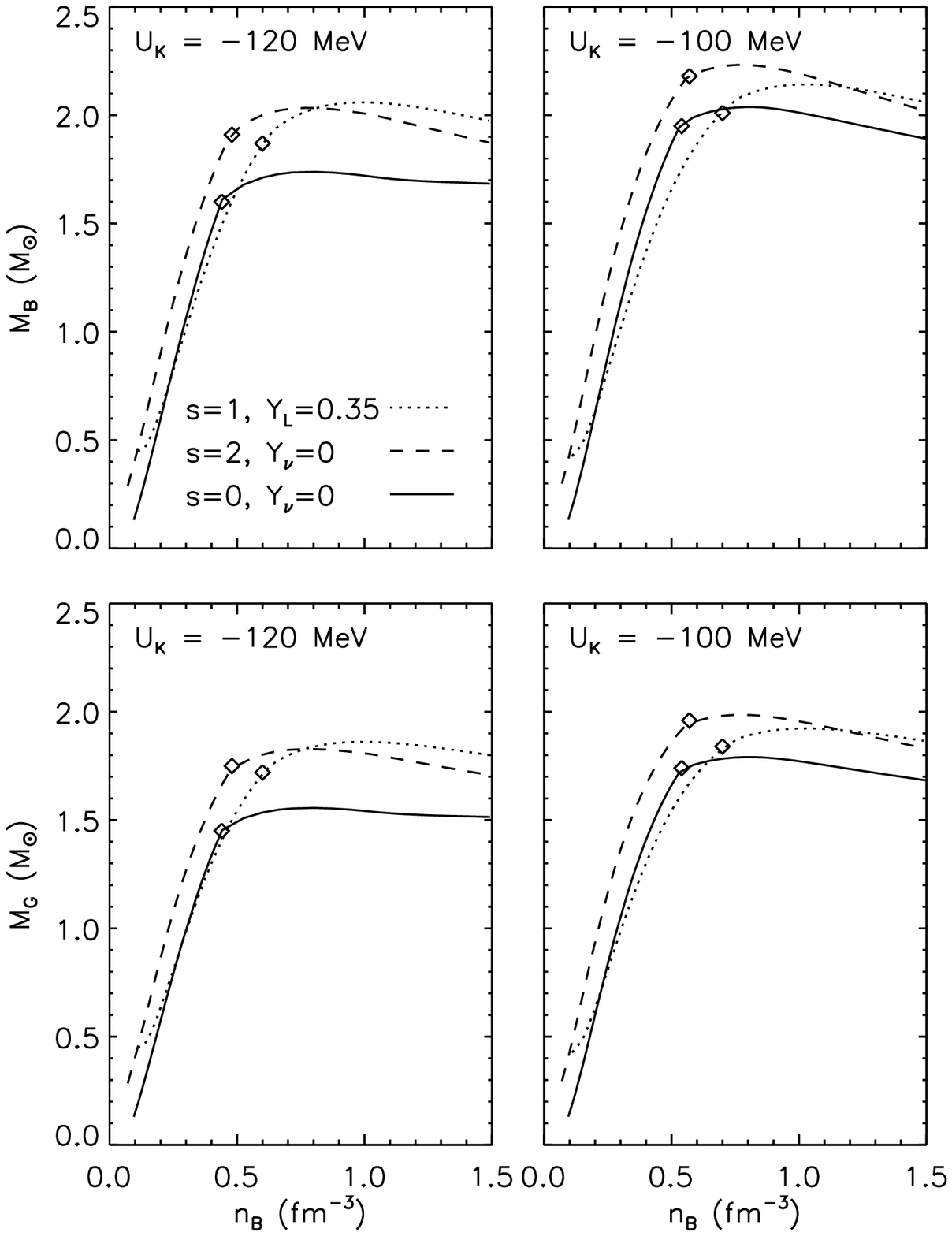}
\end{figure}

\newpage

\begin{figure}
\plotone{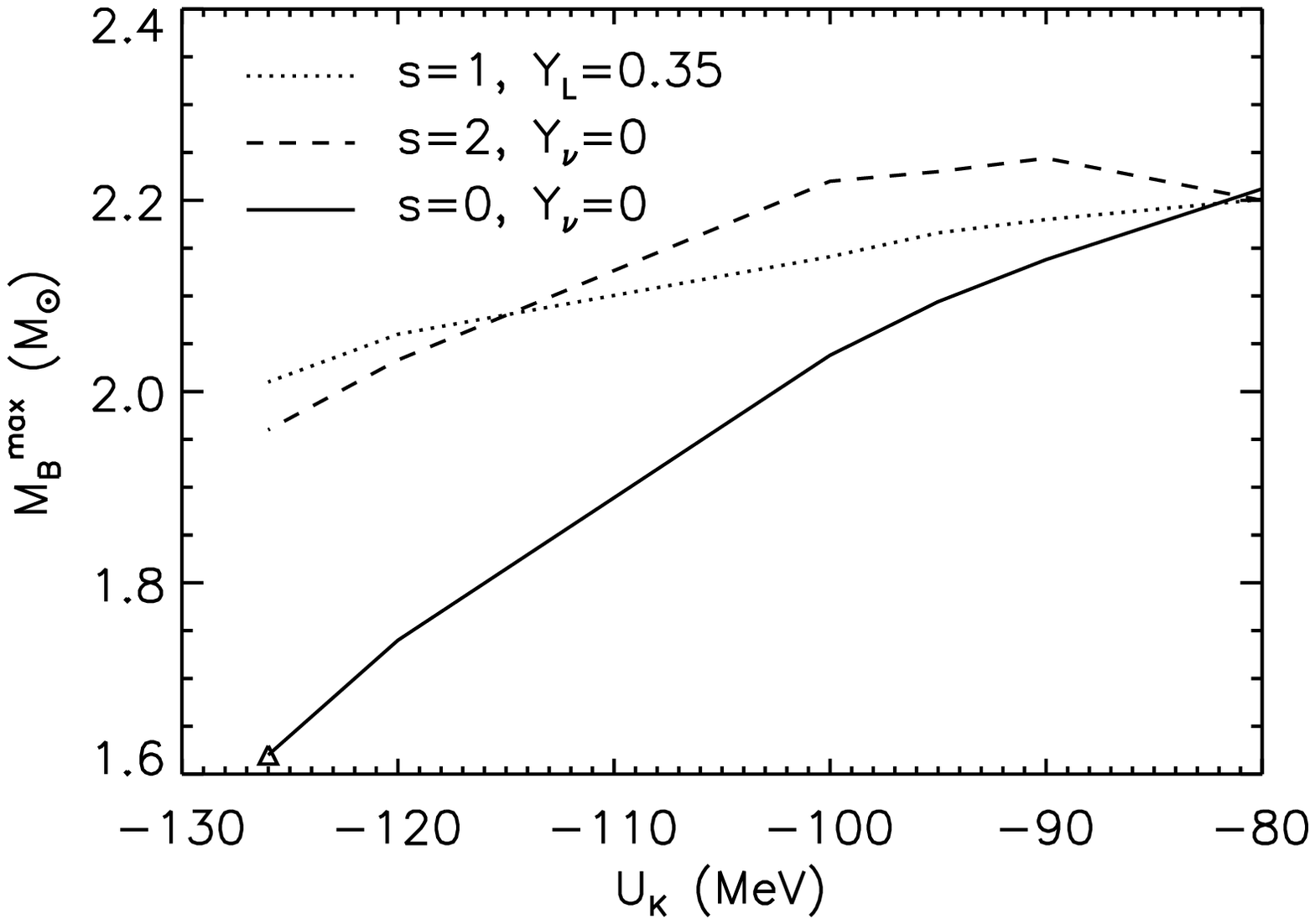}
\end{figure}

\newpage

\begin{figure}
\plotone{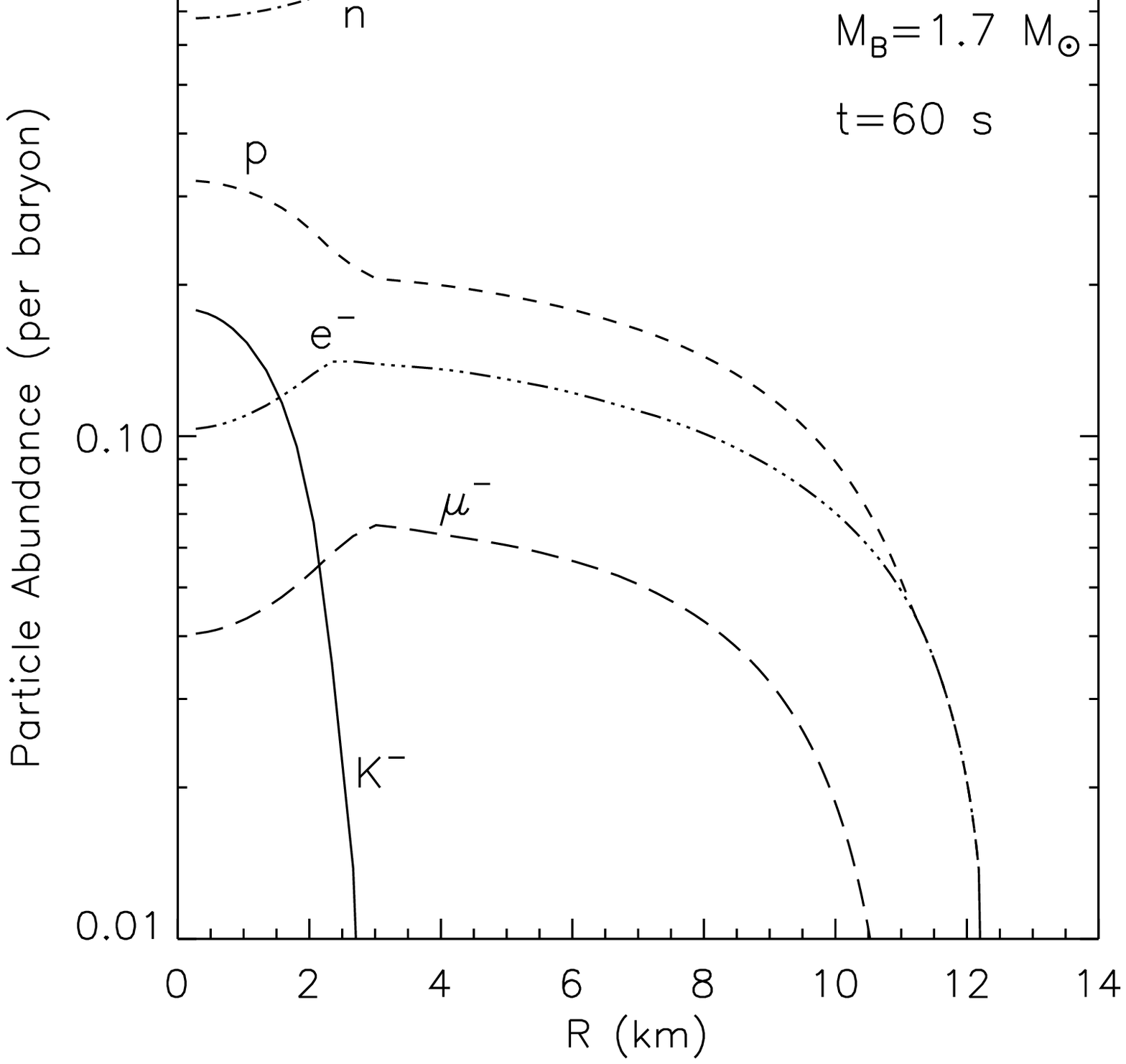}
\end{figure}

\newpage

\begin{figure}
\plotone{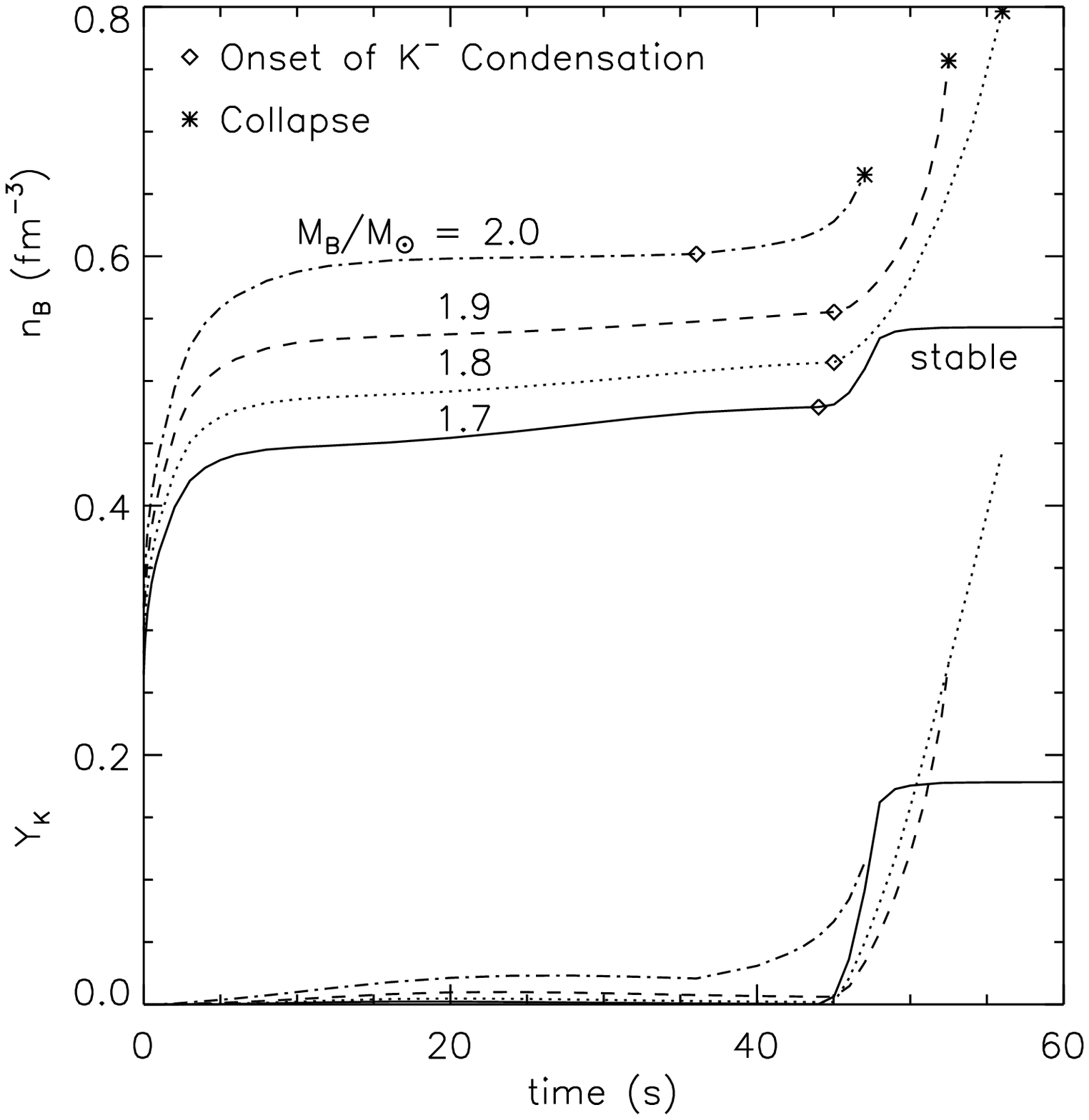}
\end{figure}

\newpage

\begin{figure}
\plotone{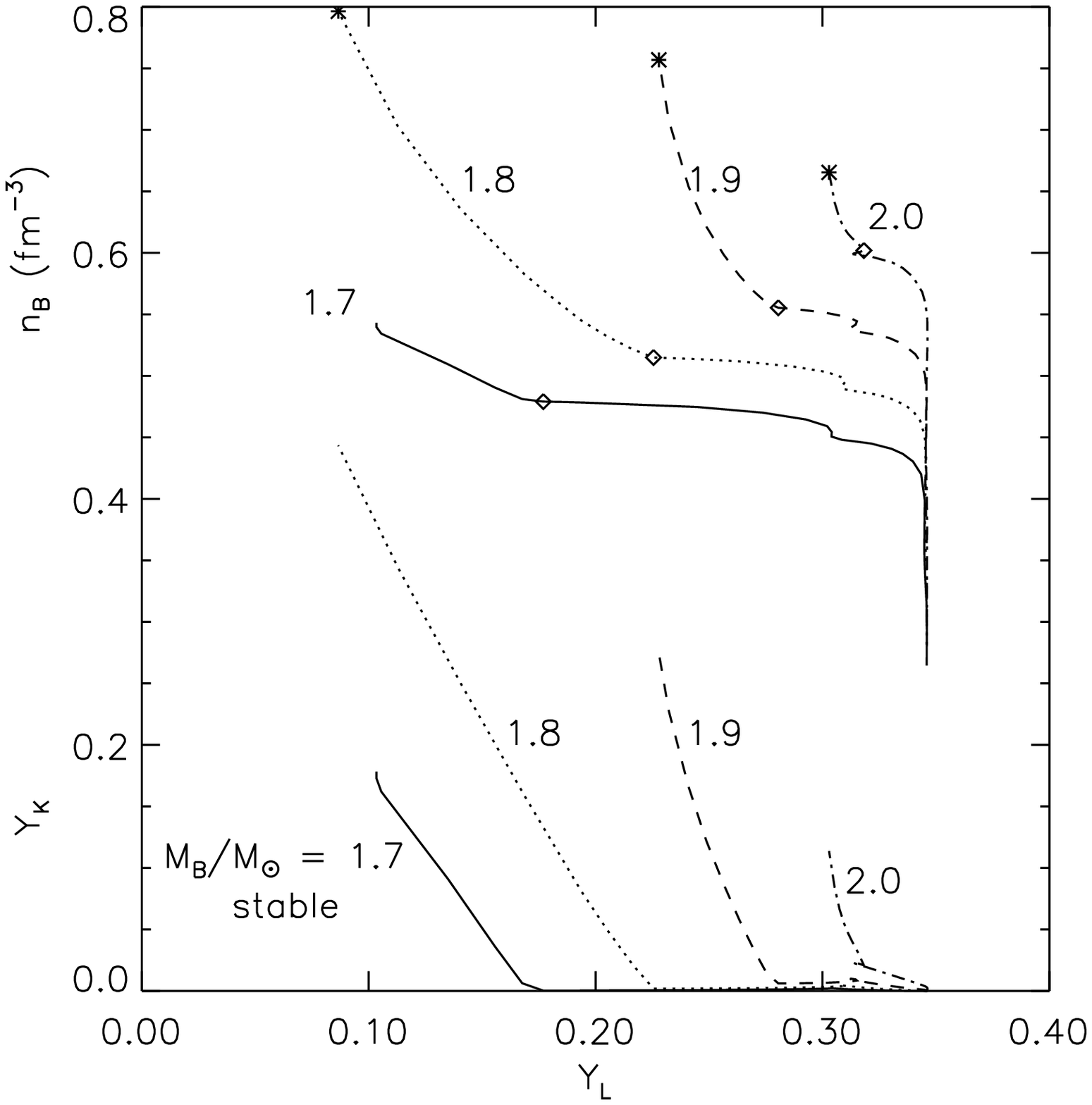}
\end{figure}

\newpage

\begin{figure}
\plotone{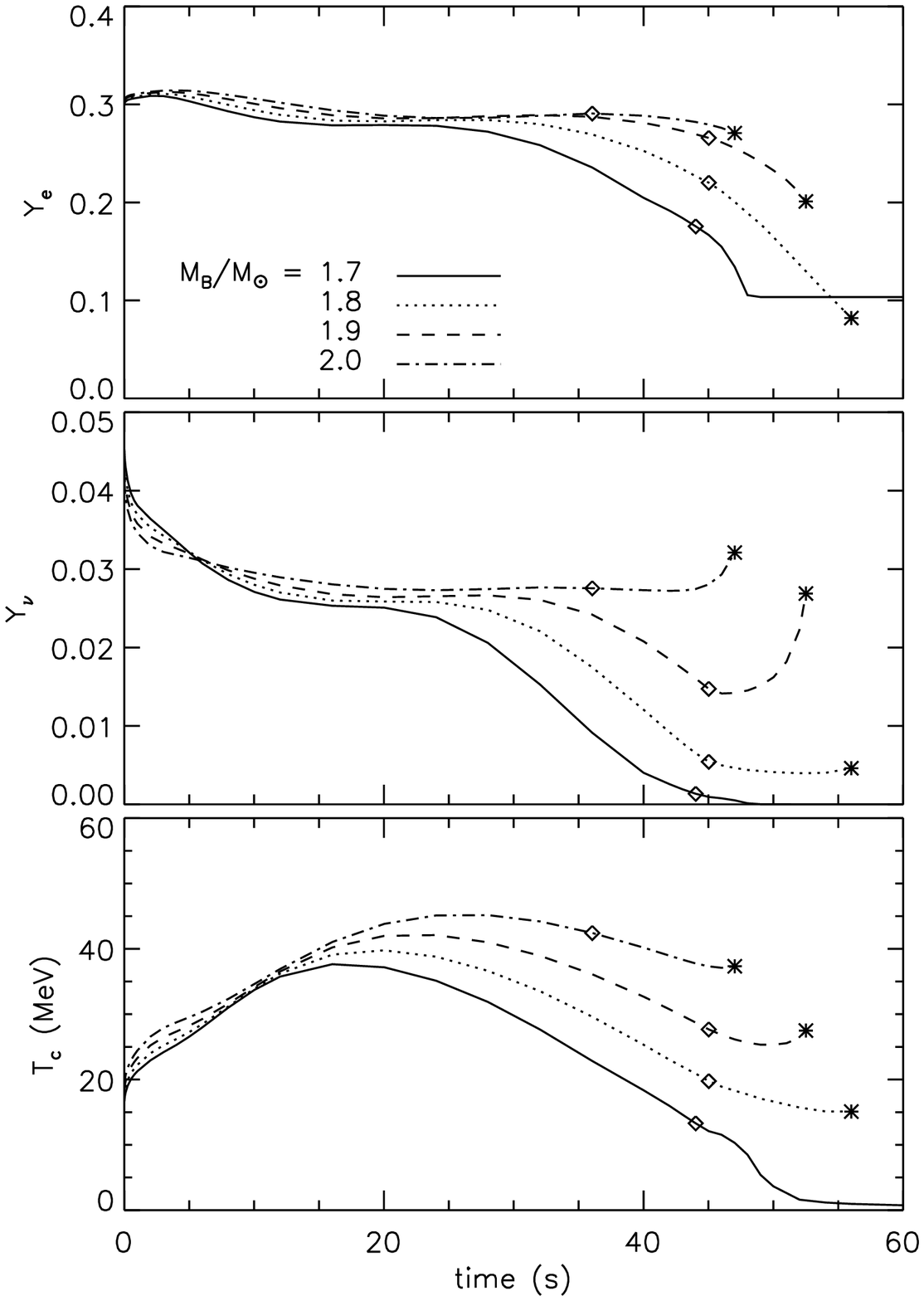}
\end{figure}

\newpage

\begin{figure}
\plotone{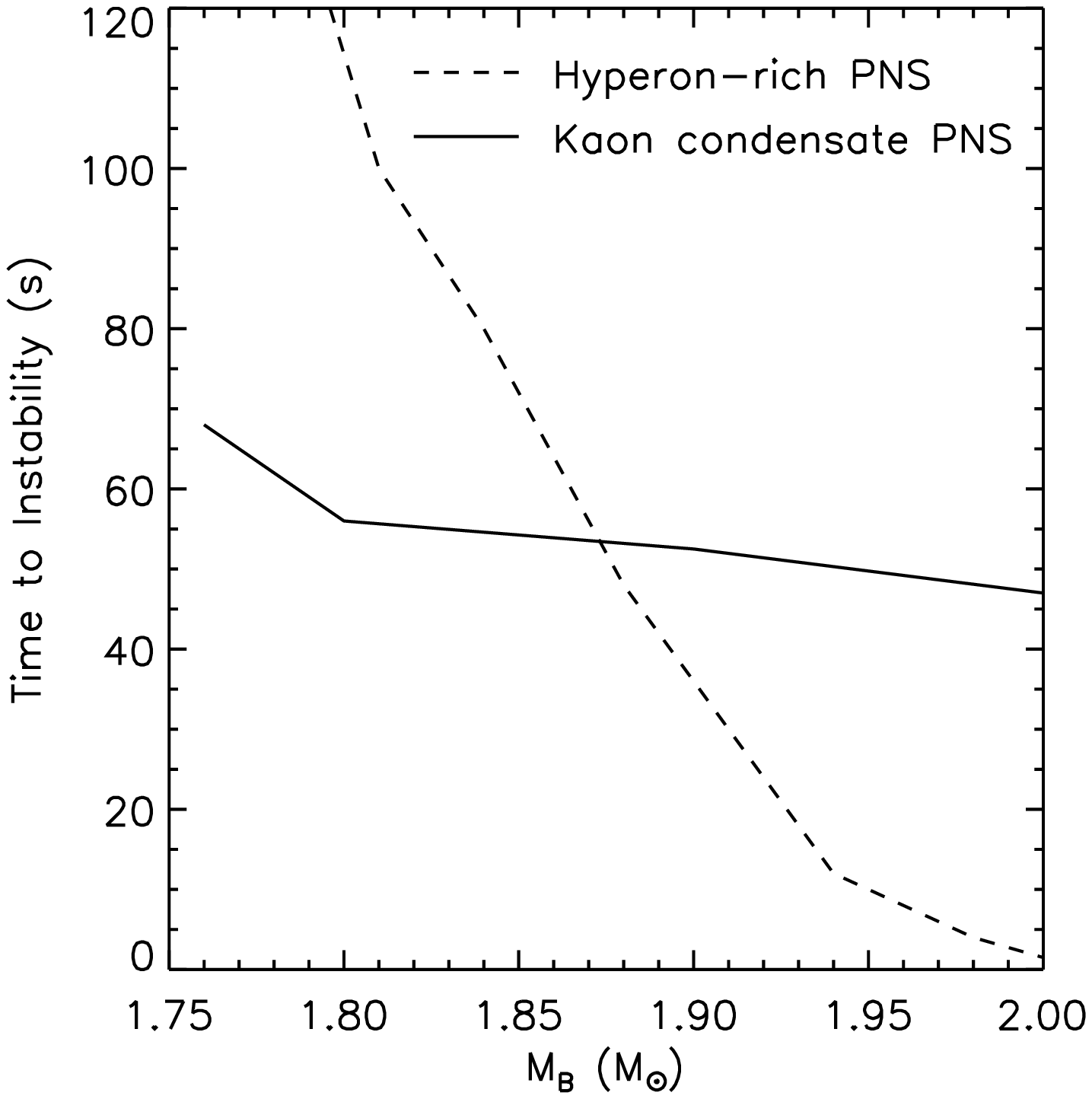}
\end{figure}

\newpage

\begin{figure}
\plotone{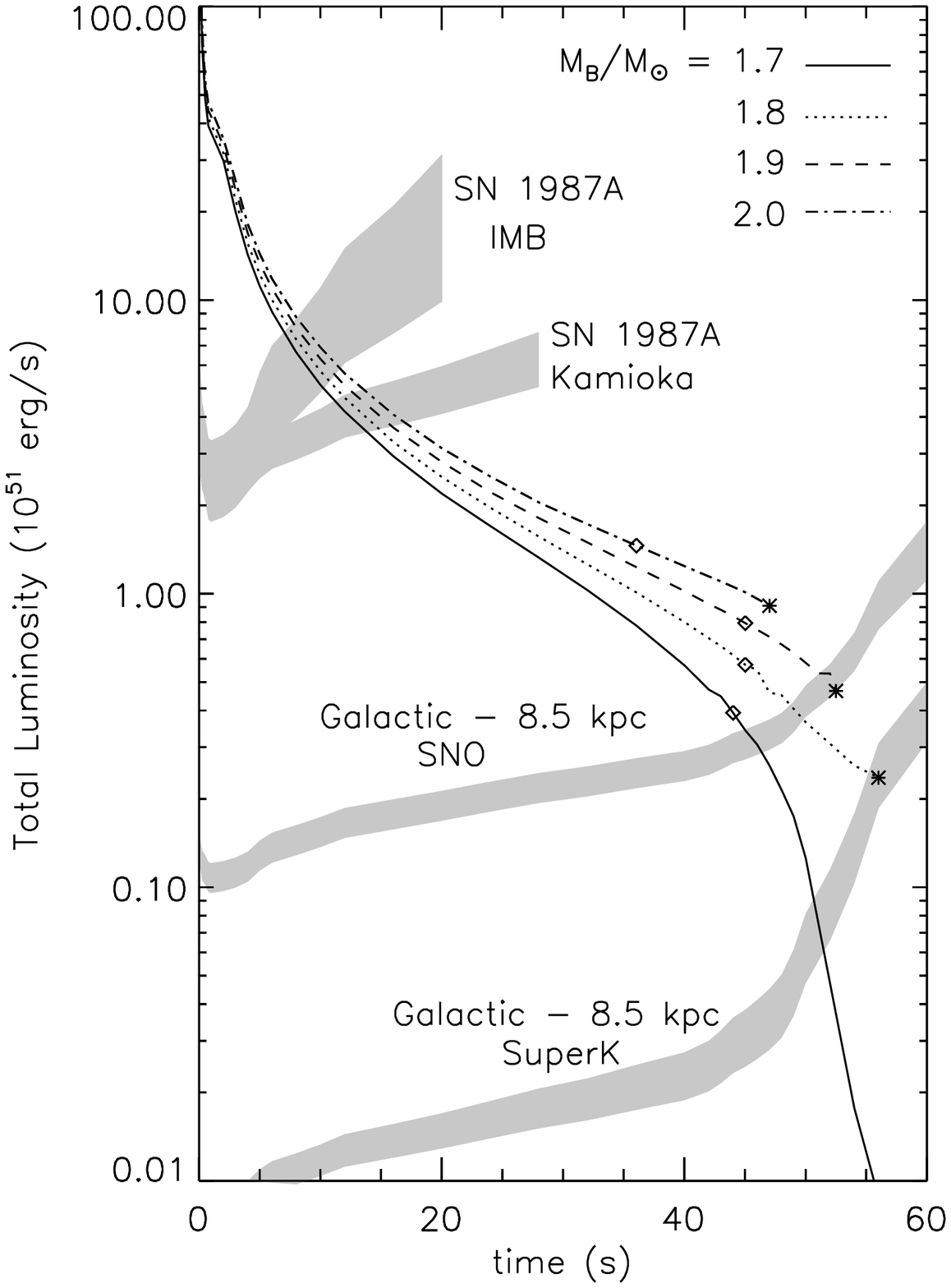}
\end{figure}

\newpage

\begin{figure}
\plotone{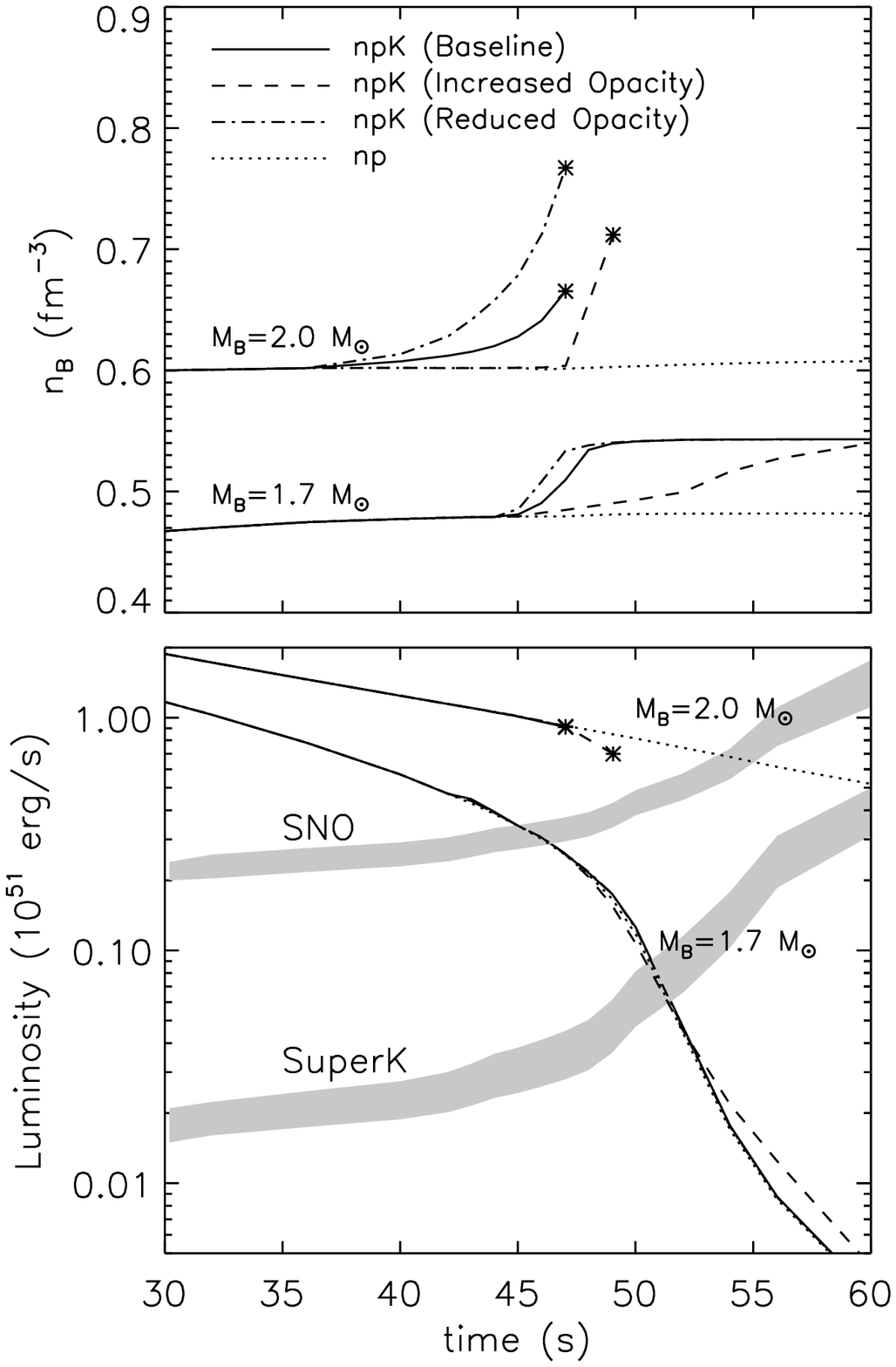}
\end{figure}

\newpage

\begin{figure}
\plotone{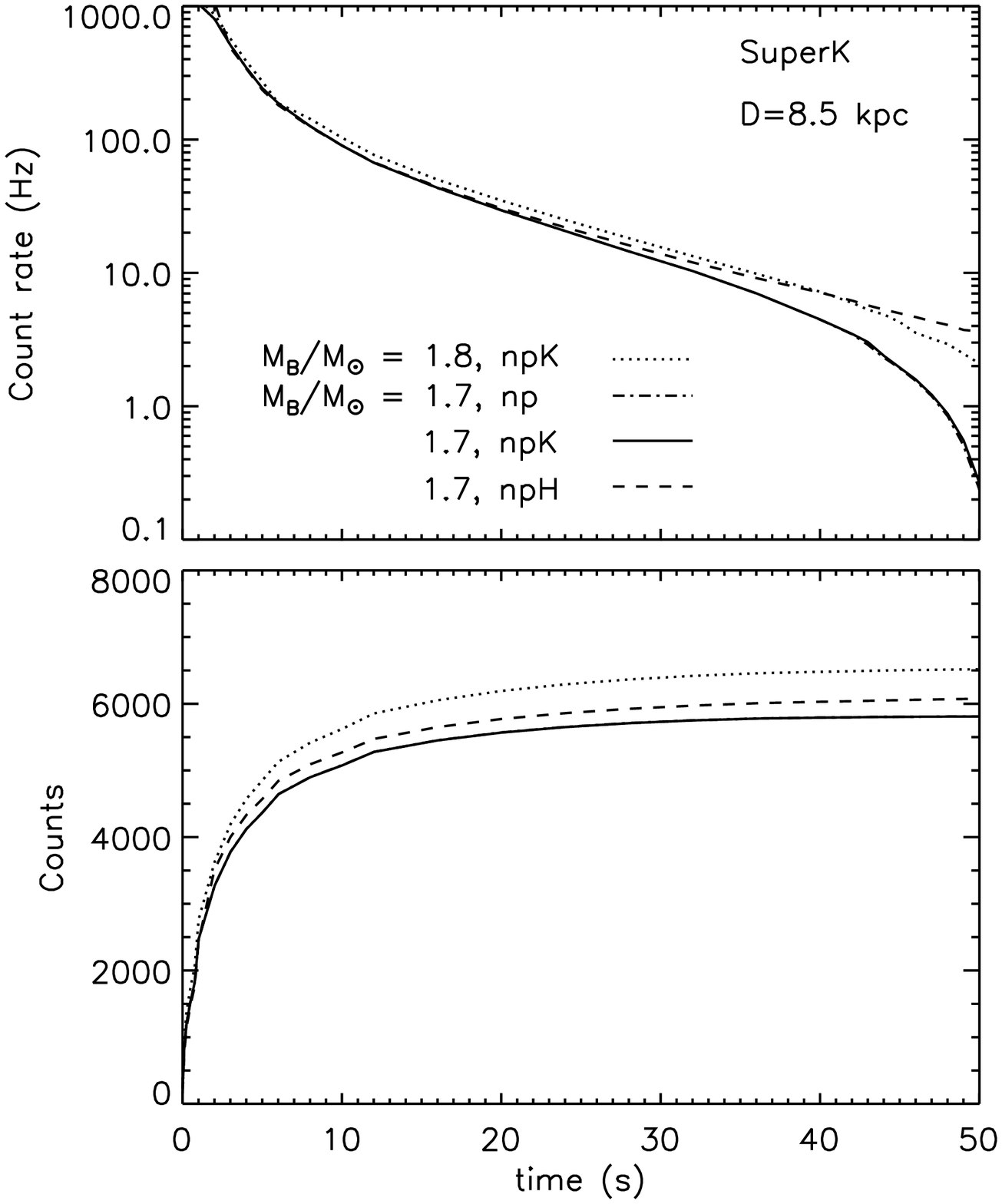}
\end{figure}

\end{document}